\documentclass[aps,floats,pre,showpacs
]{revtex4}

\newcommand{\he}{{\tilde h}}
\newcommand{\tZ}{{\tilde Z}}

\newcommand{\aU}{{\rm Ad}_{\hat U}}

\newcommand{\ppi}{{\hat {\cal \pi}}}

\usepackage{amsfonts,amsmath,mathrsfs} \usepackage{bm} \usepackage{dcolumn}
\usepackage{epsfig} \usepackage{latexsym}

\begin{document}

\title{Theory of localization and resonance phenomena in the quantum kicked rotor}
% \title{Field theory of the quantum kicked rotor}
% \title{Universal field theory of dynamical localization in quantum kicked rotors}

\author{Chushun Tian and Alexander Altland}

\affiliation{Institut f{\"u}r Theoretische Physik, Universit{\"a}t
zu K{\"o}ln, K{\"o}ln, 50937, Germany}

% \date{\today}
\begin{abstract}
 We present an analytic theory of quantum interference and Anderson
 localization in the quantum kicked rotor (QKR). The behavior of the system
is known to depend sensitively on the value of its effective
 Planck's constant $\he$. We here show that for rational values of $\he/(4\pi)=p/q$, it
 bears similarity to a disordered metallic ring of circumference $q$
 and threaded by an
 Aharonov-Bohm flux. Building on that correspondence, we obtain
 quantitative results for the time--dependent behavior of the QKR
 kinetic energy, $E(\tilde t)$ (this is an observable which sensitively probes the
 system's localization properties). For values of $q$ smaller than the localization length
 $\xi$, we obtain scaling $E(\tilde t) \sim \Delta \tilde t^2$, where
 $\Delta=2\pi/q$ is the quasi--energy level spacing on the ring. This scaling is
 indicative of a long time dynamics that is neither localized nor
 diffusive. For larger values $q\gg \xi$, the functions $E(\tilde t)\to
 \xi^2$ saturates (up to
 exponentially small corrections $\sim\exp(-q/\xi)$), thus reflecting
 essentially localized behavior.
\end{abstract}


\maketitle

\section{Introduction}
\label{intro}

The kicked rotor, or standard map~\cite{Chirikov59} is one of the most
prominent model systems of nonlinear dynamics.  In spite of its very
simple construction -- the rotor Hamiltonian depends on just a single
parameter, the dimensionless kicking strength, $K$ -- it shows complex
physical behavior. Specifically, the classical rotor undergoes a
transition from integrable to chaotic dynamics as the kicking strength
is increased. Quantization in
the chaotic regime leads to a quantum system that bears strong
similarity to a disordered multi-channel quantum wire.  Much like in
these systems, the dynamics of the quantum kicked rotor (QKR) is governed by mechanisms of
quantum interference and localization~\cite{Casati79,Chirikov81}.  The applied interest in researching these structures has
increased since the mid nineties \cite{Raizen00} when concrete
realizations of the rotor in atom optics settings became an
option~\cite{Raizen95}.

The analogies to condensed matter systems notwithstanding, there are
also a number of important differences. The QKR is not a genuinely
disordered system. Rather, its effective stochasticity roots in
mechanisms of incommensurability. The differences from generic
disorder manifest themselves in anomalies showing as Planck's
constant assumes peculiar ``resonant''
values~\cite{Casati79,Izrailev79,Chang85,Izrailev90,Casati00}. (In
this paper, we will not discuss the role of \textit{classical}
anomalies, such as accelerator mode formation \cite{Zaslavsky97} at
special values of $K$.) At these values, the system becomes
effectively periodic (in the space of angular momentum quantum
numbers) and generally escapes localization.  Various aspects of QKR
resonances have been discussed before in terms of general
observations, phenomenological reasoning and direct diagonalization
(see Ref.~\cite{Izrailev90} for review.) However, a microscopic
theory of the interplay of resonances and localization has not yet
been formulated. The construction of such a theory is the subject of
the present paper.

Below, we will introduce an analytic and parametrically controlled
theory of localization phenomena in the QKR both on and off resonant
values of Planck's constant, $\he$. It turns out that at resonant
values, $\he=4\pi p/q$, with $p,q$ coprime numbers, localization
phenomena in the QKR can be described in terms of a field theory
defined on a ring subject to an Aharonov-Bohm flux (the latter
implementing the Bloch phases pervasive in the physics of
periodically extended systems.) We will apply perturbative and
non--perturbative methods to analyze this field theory in terms of
different observables. Emphasis will be put on the discussion of the
time dependent increase in the rotor's kinetic energy, an observable
that carries immediate information on the transport characteristics
of the system.

Before turning to a qualitative discussion of  results, it is
worthwhile to put the present theory into the context of earlier
work. The connections to disordered systems were first discussed in
Ref.~\cite{Fishman83}. That work introduced a mapping of the QKR
Hamiltonian onto an effectively disordered tight binding system.
However, the long-ranged correlated 'disorder' inherent to that
model made it difficult to treat analytically. A different approach
was introduced in Ref.~\cite{Altland93}, where a diagrammatic
approach conceptually similar to the impurity diagram language of
disordered systems was introduced. In this work, perturbative
quantum interference processes (``weak localization'') were
discussed in terms of low energy fluctuations in angular momentum
space (for a refined implementation of the diagrammatic approach,
see Ref.~\cite{Tian05}.) These results inspired a subsequent field
theoretical study~\cite{Zirnbauer96a} in which the large scale
dynamics of the quantum kicked rotor was mapped onto a
one-dimensional supersymmetric nonlinear $\sigma$--model (otherwise
known as the effective field theory of Anderson localization in
disordered quasi-one-dimensional wires \cite{Efetov97}.)

None of these works payed attention to the resonance phenomena at
rational $\he/(4\pi)$, a lack of resolution
that has prompted some criticism~\cite{Casati98,Casati00}.
Specifically, it has been argued that the insensitivity of earlier
work to the algebraic characteristics of $\tilde h$ might be
indicative of ad hoc assumptions hidden in the construction. One of
the motivations behind the present work is to show that these objections
have been ill founded. Rather, we will see that the
field theory below -- by and large a refined variant of the earlier
theory~\cite{Zirnbauer96a} -- is capable of describing the system
from large scales down to the fine structure of individual
quasi-energy levels.

Nonetheless, it is worthwhile to point out that our theory is far from
'mathematically rigorous'.
Rather, it is based on a mapping of  QKR correlation functions onto
a functional integral (exact) followed by an identification of
 field configurations of least action, plus an integration over
fluctuations. The last two steps are approximate, if parametrically
controlled (by a small parameter $\tilde h/K$.)
In this sense, our theory below falls into the general framework of
semiclassical approaches to nonlinear systems. The semiclassical limit
enables us to explicitly describe system observables which more
rigorous lines of reasoning~\cite{Jitomirskaya03} can only
characterize in implicit terms. We also emphasize that no averaging
procedures other than a global average over the quasi-energy
spectrum will be performed throughout.

The rest of this paper is organized as follows: we start with a
qualitative discussion of the system and a summary of our main
results (Section~\ref{sec:dynam-local}). In
Section~\ref{sec:functional5} we introduce the functional integral
approach to the problem, which will be reduced to an effective
semiclassical action in Section~\ref{sec:sigma}. The low energy
physics of this action will then be discussed in Sections \ref{off
resonancs} (off-resonance) and \ref{resonances} (on resonance). We
summarize in Section \ref{sec:conclusion}. Some technical details
are deferred to Appendices
\ref{sec:derivationSfl}-\ref{sec:mass-mode-fluct}.

\section{Qualitative discussions}
\label{sec:dynam-local}

The QKR is a periodically driven (``kicked'') quantum particle
moving on a circle. In units where the particle mass and the
circular radius are set to unity, its Hamiltonian reads
\begin{eqnarray}
% i\hbar \frac{\partial }{\partial t}|\psi(t)\rangle &=& {\hat H}
% (t)|\psi(t)\rangle \,, \label{dynamics}\\
{\hat H} (t) &=& \frac{{\hat l}^2}{2}+k\cos{\hat \theta}\,
\sum_m\delta (t-mT), \label{Hamiltonian}
\end{eqnarray}
where $T$ and $k$ are the period and the kicking strength,
respectively. The angular operator, $\hat \theta$, and the angular
momentum operator, $\hat l$ obey the canonical quantization
condition $[{\hat \theta}\,, {\hat l}]=i\hbar$ where ${\hat
l}=i\hbar\frac{\partial}{\partial\theta}$\,. In rescaled
dimensionless time, $t\rightarrow t/T\equiv \tilde t$, the dynamics
of the system is determined by two parameters: the effective
Planck's constant $\he=\hbar T$ and the classical nonlinear
parameter $K=kT$. (In the chosen units, both parameters are
dimensionless.)

\subsection{Symmetries}
\label{sec:symmetries}

At integer times $\tilde t = 0,1,2,\dots$, the solution of the
Schr\"odinger equation $i\he \,\partial_{\tilde t} |\psi(\tilde
t)\rangle = T^2 \hat H |\psi(\tilde t)\rangle$ can be expressed as
$|\psi({\tilde t})\rangle = {\hat U}^{\tilde t} |\psi(0)\rangle$,
where $\hat U$ is the Floquet operator, i.e. the unitary operator
governing the time evolution during one elementary time step:
\begin{equation}
{\hat U}=\exp\left(\frac{i\tilde h {\hat
n}^2}{4}\right)\exp\left(\frac{iK\cos{\hat
\theta}}{\he}\right)\exp\left(\frac{i\tilde h {\hat n}^2}{4}\right),
\label{U}
\end{equation}
where we have introduced $\hat n \equiv \hat l /\hbar$ as a rescaled
angular momentum operator with
integer--valued spectrum quantum.

For Planck's constants commensurable with $4\pi$, $\he=4\pi p/q$,
with coprime $p,\,q\in \Bbb{N}$, the Floquet operator is invariant
under the shift ${\hat n}\rightarrow {\hat n}+ q$, or
\begin{align}
\label{Tq_sym}
[\hat U, \hat T_q]=0,
\end{align}
where $\hat T_q$ is the translation operator by $q$ steps, $\hat T_q
|n\rangle = |n+q\rangle$ on eigenstates $\hat n|n\rangle = n
|n\rangle$.
  This discrete translational
symmetry implies that the (quasi)energy states of the Floquet
operator can be expanded in a basis of Bloch states,
\begin{align*}
  e^{i\theta n} \psi_{n,\theta},
\end{align*}
where $\theta\in [0,2\pi/q]$ and
$\psi_{n+q,\theta}=\psi_{n,\theta}$. The infinite extension of Bloch
states in angular momentum space means the absence of localization
at these ``resonant values'' of $\tilde
h$.~\cite{Casati79,Izrailev90,Chang85,Izrailev79,Casati00} (However,
for periodicity intervals $q$ much larger than the intrinsic
localization length of the system,
% $\xi$,
the wave function amplitudes
% $\psi_\phi$
at the boundaries of the ``unit cell'' are exponentially small,
which means that the system behaves effectively localized.) Below,
we will explore the interplay of localization and Bloch periodicity
in some detail. Our  discussion will include the case of
\textit{irrational} $\tilde h/4\pi$ as a limit of co--prime values
$p/q$ with diverging $q$. We will assume that $q\gg K/\tilde h
\equiv \ell$, which means that the angular momentum unit cell is
larger than the length scales at which diffusive dynamics begins to
form.

In dirty metals, the manifestations of localization depend sensitively
on the presence or absence of time reversal, $t\rightarrow -t, r
\rightarrow r, p \rightarrow -p$ (where $r$ and $p$ are coordinate and
momentum respectively.) In the context of the kicked rotor, momentum
and coordinate space change their roles, which is why time reversal
followed by space inversion, ${\rm T}_c:\, t \rightarrow -t, \theta
\rightarrow -\theta , l \rightarrow l$ becomes a relevant symmetry. It
has been shown~\cite{Altland93,Tian05,Smilansky92} that ${\rm T}_c$
plays a role analogous to time reversal in metals. The Hamiltonian
\eqref{Hamiltonian} is ${\rm T}_c$--invariant, and in our analysis
below, we need to take this symmetry into account.
($\mathrm{T}_c$--invariance may be broken e.g. by a shift $\cos(\hat
\theta) \to \cos(\hat \theta+a), \;\; a=\mathrm{const.}$ However, in
this paper, we focus on the invariant system.) In a basis of angular
momentum eigenstates, $\{|n\rangle\}$,\; $\hat l|n\rangle = \he n\,
|n\rangle,\;n\in \Bbb{Z}$\,, the symmetry ${\rm T}_c$ acts by matrix transposition
$A_{nn'}\to (A^T)_{nn'}\equiv A_{n'n}$ (much like conventional time
reversal acts by transposition in a real space basis.)

\subsection{Off resonance}
\label{sec:phen-local}

Let us temporarily assume that $\tilde h \in 4\pi \Bbb{R}\backslash
\Bbb{Q}$ is irrational (or rational with $q\gg \xi$, so that the
periodicity of the system plays no role.) Throughout this paper, we
consider the system prepared in a pure initial state with zero
angular momentum, i.e., $|0\rangle \langle 0|$. Under the Floquet
dynamics it will evolve into $(\hat U)^{\tilde t}|0\rangle \langle
0| (\hat U^\dagger)^{\tilde t}$. Its (squared) deviation from the
point of departure is measured by the expectation value $E(\tilde
t)\equiv {1\over 2}\mathrm{tr}(\hat n^2 (\hat U)^{\tilde t}|0\rangle
\langle 0| (\hat U^\dagger)^{\tilde t})$. Comparison with the
Hamiltonian \eqref{Hamiltonian} shows that this quantity may be
interpreted in terms of the system's change in kinetic energy.  In
the classical limit, $E(\tilde t)$ value grows indefinitely in time,
a phenomenon that may be interpreted as a mechanism of 'heating'.
More precisely,
$$
E(\tilde t) \sim D_0 \tilde t,
$$
where $D_0\equiv (K/2\tilde h)^2$ has the status of a diffusion
constant in angular momentum space. Notice that the energy
expectation value can be represented as $E({\tilde
t})=\frac{1}{2}\sum_n n^2K(n,0;\tilde t)$, where $K(n,n';t)$ is the
density-density correlation function in angular momentum space (for
a precise definition, see Section \ref{sec:functional5} below.) The
linear growth in time is indicative of diffusive behavior
$K(q,\omega)\sim (-i\omega +D_0 q^2)^{-1}$, where $(\omega,q)$ is
Fourier conjugate to $(\tilde t,n)$. Empirically, it has long been
known \cite{Chirikov81} that in the \textit{quantum} model, the
linear growth in energy comes to an end at times $\tilde t_L\sim
D_0$. Simple scaling arguments show that this saturation implies a
crossover $D_0\to D(\omega) \sim i\omega$ at frequencies
$\omega\lesssim D_0^{-1}$. The scaling $D(\omega) \sim i\omega$ is a
manifestation of quantum localization~\cite{Efetov97}.

A microscopic quantum theory of the system must be capable
of establishing diffusive dynamics at intermediate time scales, and to
describe the quantum interference processes rendering the diffusion
constant frequency dependent.

\subsection{On resonance}
\label{sec:resonance}

\begin{figure}[h]
  \centering
  \includegraphics[width=10.5cm]{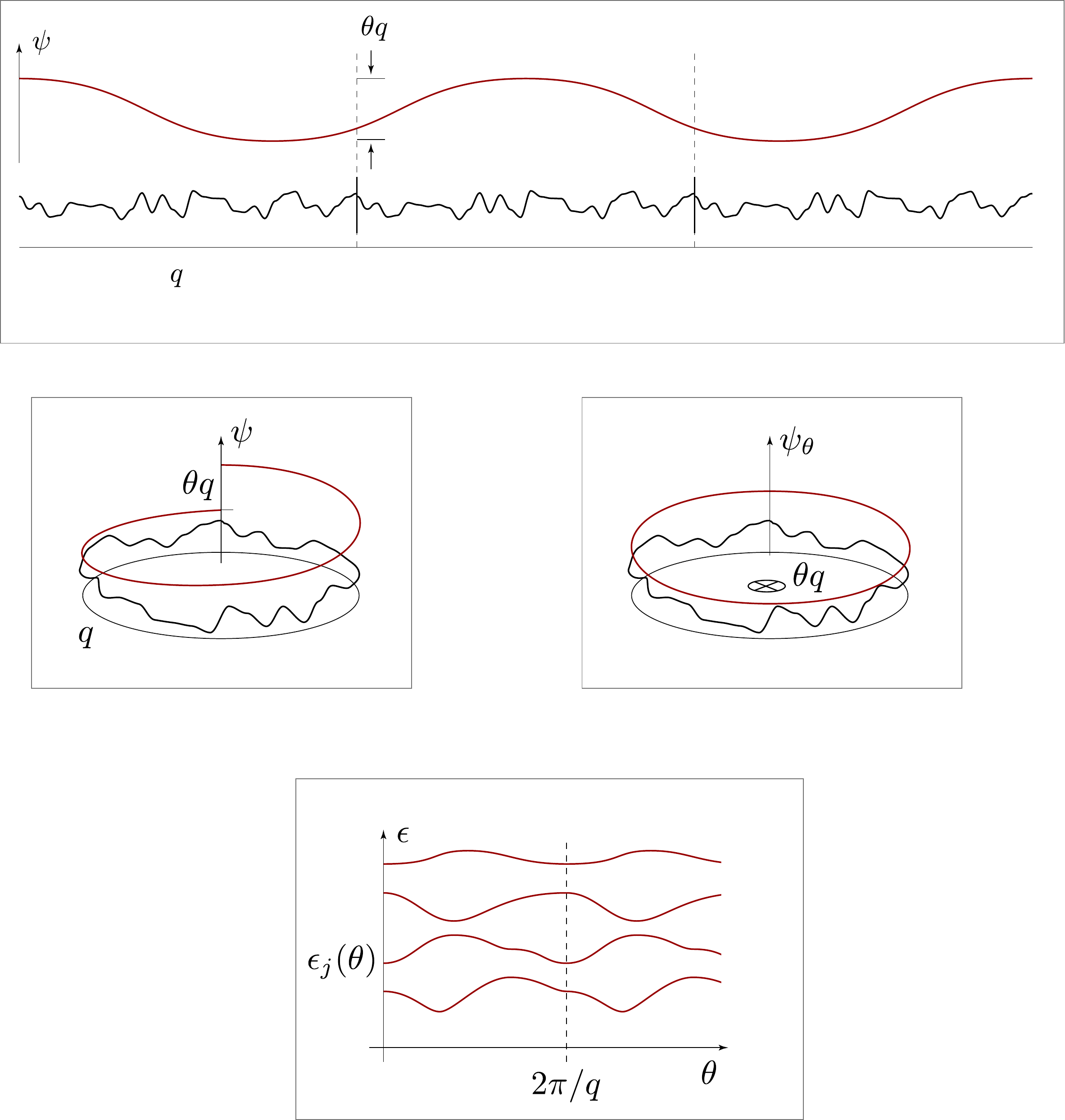}
  \caption{Wave functions and spectrum of the periodically extended system.}
  \label{fig:periodic}
\end{figure}

We now discuss what happens for rational values $\tilde h = 4\pi
p/q$. The commutativity \eqref{Tq_sym} means that we are considering
a variant of a $q$-periodic quantum system. For the convenience of
the reader, a few basic structures of periodically extended quantum
systems are recapitulated in Fig.~\ref{fig:periodic}. The Bloch
function basis can be understood as a basis whose elements exhibit a
definite phase change $\theta q\in [0,2\pi]$ across a single unit
cell (top panel.) For a given $\theta$, one may then consider the
system within the reduced scheme of just a single unit cell, subject
to twisted boundary conditions $\psi(0) = \psi(q)\exp(i\theta q)$
(middle left). Equivalently, one may interpret the system in terms
of an Aharonov-Bohm ring (middle right) subject to a gauge flux
$\theta q$. The $q$ energy levels $\epsilon_j(\theta)$ of this
system (bottom) form a $2\pi/q$ periodic family, each $j$ defines a
Bloch band.

Pioneering work on the physics of the kicked rotor at these
``resonant'' values of Planck's constant has been done by Izrailev
and Shepelyansky (IS)~\cite{Izrailev79} (see Ref.~\cite{Izrailev90}
and Section \ref{sec:theory-izra-shep} below for review.) In this
work, it has been argued that at resonance, the energy stored in the
system increases as
\begin{align}
  \label{Eincrease}
  E(\tilde t) = \eta \tilde t^2 + a \tilde t + b(\tilde t),
\end{align}
(designation of constants $\eta,a$ and function $b(\tilde t)$ taken
from the review Ref.~\cite{Izrailev90}). Let us briefly discuss the
meaning of this expansion, as discussed in IS. Eq. \eqref{Eincrease}
has the status of a large time asymptotic. For times large enough
such that $\tilde t^{-1}$ is smaller than the level spacing of the
unit cell, $\Delta=2\pi/q$, we are in a ``deep quantum regime'',
where transitions between individual Bloch bands can be neglected.
In this regime, the energy increase is governed by expressions such
as (symbolic notation)
 \begin{align*}
   E(\tilde t) \sim \int d\theta d\theta'
   \,e^{i(\epsilon_j(\theta)-\epsilon_j(\theta'))\tilde t - i
     n(\theta-\theta')}n^2\sim \tilde t^2.
 \end{align*}
Thus, the $\tilde t^2$ contribution is a measure of the Bloch--phase
auto--correlation of individual levels. The auto--correlation of
levels also yields a  quantum correction linear in $\tilde t$, and a
residual contribution $b(\tilde t)$ of indefinite time dependence.
The coefficients of both terms, $\eta$ and $a$, reflect the
dispersion of Bloch bands, and vanish if bands are flat. Finally, in
regimes with localization, $\xi\ll q$ we expect the increase in
$n^2$ to come to an end at large times $\tilde t \sim \xi^2/D_0$. At
larger times, $E(\tilde t)$ will saturate. (With exponentially small
corrections $\sim \exp(-q/\xi)$ proportional to the wave function
overlap across the ring.)

The construction of IS that led to \eqref{Eincrease} has been formal
in that it relied on explicit knowledge of eigenfunctions and
-values of the Floquet operator; apart from a few exceptional values
of $q$, no concrete results  for the coefficients of the expansion
have been available. Equally important, at times shorter than
$\Delta^{-1}$ mechanisms of energy diffusion different from those
discussed in IS play a role. Specifically, for small values of time,
the finite extension of the unit cell is not yet felt, and angular
momentum will diffuse as in an infinite system. This generates a
(leading) $\sim t$ dependence in $b(\tilde t)$, which is purely {\it
  classical}. (Within the framework of a quantum approach, diffusion
has to do with inter--band transitions.) At short times, the profile
of $E(\tilde t)$ will, thus, be different from \eqref{Eincrease}.

Below, we will analytically derive the function $E(\tilde t)$ for
periodicity volumes obeying $\ell<q<\xi$. As a result, we obtain a
universal scaling function
\begin{align}
  \label{Et_scaling}
  E(\tilde t) &= D_q q \,F\left({\tilde t\over q}\right),\cr
F(x)&=  \left\{ \begin{array}{cl}
                       x + {x^3\over 3},&\quad (qE_c)^{-1}\ll x<1 \\
                       x^2 + \frac{1}{3}
                       , &\quad x>1
                     \end{array}\right.,
\end{align}
where $E_c = D_q/q^2$ is the inverse of the classical diffusion time
through the periodicity volume. Referring for a more substantial
discussion to Section \ref{sec:reson-regime-lqxi}, we here merely
note that Eq. \eqref{Et_scaling} predicts diffusive scaling $E\sim
D_q \tilde t$ for times $\tilde t<\Delta^{-1}$. For larger times, we
obtain a quadratic increase, $E\sim \eta \tilde t^2$, governed by
the universal coefficient $\eta=D_q/q$. For the discussion of the
meaning of this coefficient and of the correction terms in
\eqref{Et_scaling}, we refer to section \ref{sec:reson-regime-lqxi}.

In Section \ref{sec:reson-regime-gxi}, we will briefly discuss the
regime of large resonance volumes $q>\xi$. Our results are in line
with the general expectations formulated above.

\subsection{Close to fundamental resonance}
\label{sec:other-symmetries}

The sensitivity of the rotor to the algebraic properties of $\tilde
h$ manifests itself not only on resonances, but also in the close
vicinity of low--order resonant values. To be specific, consider
values $\tilde h = 4\pi p + \epsilon$, close to the fundamental
resonance $q=1$. Straightforward substitution into \eqref{U} shows
that the Floquet operator defined for the pair $(\tilde h,K)$, is
identical to one with $(\tilde h_\epsilon,K_\epsilon)$,
where~\cite{Fishman03}
\begin{eqnarray}
\label{eps_class}
  \tilde h_\epsilon\equiv \epsilon,\qquad K_\epsilon \equiv {K\epsilon
    \over 4\pi p +\epsilon}.
\end{eqnarray}
This duality phenomenon has been dubbed ``$\epsilon$-classics'': a
rotor with large value of Planck's constant becomes equivalent to
one with small $\tilde h\sim \epsilon$. Faithfulness to this
symmetry is a benchmark criterion which the present approach obeys.
In practice, this means that system characteristics such as the
localization length will depend on $\tilde h$ and $K$ through
certain invariant combinations (that have been conjectured
phenomenologically before \cite{Shepelyansky87a}.)

We now turn to the quantitative formulation of our theory.

\section{Functional integral formulation}
\label{sec:functional5}

As in condensed matter physics, localization properties of the QKR
can be conveniently probed in terms of ``two particle Green
functions''. In this paper,  emphasis will be put on the density
correlation function
\begin{equation}
K_\omega(n_1,n_2)\equiv \sum_{\tilde t,\tilde t'=0}^{+\infty}
\left\langle\!\langle n_1 |({\hat U}e^{i\omega_+})^{\tilde t}|n_2
\rangle \langle n_2 |({\hat U}e^{i\omega_- })^{-\tilde t'}|n_1
\rangle\!\right\rangle_{\omega_0}, \label{correlatordefinition}
\end{equation}
%\begin{align}
%&K_\omega(n_1,n_2)\equiv\cr &\quad \sum_{\tilde t,\tilde
%t'=0}^{+\infty} \left\langle\!\langle n_1 |({\hat
%U}e^{i\omega_+})^{\tilde t}|n_2 \rangle \langle n_2 |({\hat
%U}e^{i\omega_- })^{-\tilde t'}|n_1
%\rangle\!\right\rangle_{\omega_0}, \label{correlatordefinition}
%\end{align}
where the frequency arguments, $\omega_\pm \equiv \omega_0 \pm
\frac{\omega}{2}$, with $\omega$ understood as
$\omega+i\delta,\delta\searrow 0$,
% $\omega_\pm \equiv \omega_0 \pm
% (\frac{\omega}{2}+i\delta)$ with $\delta\searrow 0$,
and we have
introduced an average over the quasienergy spectrum,
$\langle\cdot\rangle_{\omega_0}:=\int_0^{2\pi}\frac{d\omega_0}{2\pi}(\cdot)$\,.
Summation over $\tilde t,\tilde t'$ obtains
\begin{align}
K_{\omega} (n_1,n_2) =\left\langle\!\langle n_1|{\hat
G}^+(\omega_+)|n_2\rangle \langle n_2|{\hat
G}^-(\omega_-)|n_1\rangle\!\right\rangle_{\omega_0},
% \cr
\label{correlator}
\end{align}
where
\begin{equation}
{\hat G}^\pm(\omega_\pm)=\sum_{\tilde t=0}^{\infty}
\left(e^{i\omega_\pm}{\hat U}\right)^{\pm \tilde t} =
\frac{1}{1-(e^{i\omega_\pm}{\hat U})^{\pm 1}} \,, \label{Green}
\end{equation}
and the $+$ ($-$) superscript designates the retarded (advanced)
Green function. Physically, the function $K_\omega(n_1,n_2)$
describes the probability of propagation $n_2\to n_1$ in time $\sim
\omega^{-1}$. Exponential decay of $(-i\omega K_\omega)$ at
$\omega\rightarrow 0$ signals the onset of localization. The density
correlation function also contains information on the time dependent
expectation value of the energy. Comparing with the definition given
in the beginning of Section \ref{sec:phen-local}, it is
straightforward to show
\begin{align}
  \label{Ediff_Kn}
 %  E(\tilde t) ={1\over 2} \sum_n n^2 \int {d\omega\over 2\pi} (e^{-i\omega
%     \tilde t}-1) K_\omega(n,0).
 E(\tilde t) ={1\over 2} \sum_n n^2 \int {d\omega\over 2\pi} e^{-i\omega
    \tilde t} K_\omega(n,0).
\end{align}
%Here, the initial energy is subtracted so that $E(0)\equiv 0$.

In the present formalism, the consequences of time reversal symmetry
are best exposed by ``taking the square root'' of the Floquet
operator,
$$
\hat U^{\pm 1} = \hat V^\pm \hat V^{\pm T},
$$
where
\begin{equation}
{\hat V}^\pm \equiv \exp\left(\pm \frac{i\he{\hat
n}^2}{4}\right)\exp\left(\pm\frac{iK\cos {\hat \theta}}{2\he}\right)
\,. \label{Vsplit}
\end{equation}
We next introduce the matrix Green function
\begin{eqnarray}
{\tilde G}^\pm  \equiv \left(
                                      \begin{array}{cc}
                                        1 & e^{\pm \frac{i\omega_\pm}{2}} {\hat V}^{\pm} \\
                                        e^{\pm \frac{i\omega_\pm}{2}} {\hat V}^{\pm T} & 1 \\
                                      \end{array}
                                    \right)^{-1}\,.
 \label{GreenTspace}
\end{eqnarray}
We will refer to the newly introduced two-component structure as
``time reversal space'', or ``T''--space and label it by indices $t,t',\dots$.
The matrix $\tilde G$ is related to the original Green function as
$
\hat G=\tilde G_{11}$, i.e.
\begin{align}
K_{\omega} (n_1,n_2) =\left\langle\!\langle n_1|\tilde G^+_{11}|n_2\rangle \langle n_2|\tilde G^-_{11}|n_1\rangle\!\right\rangle_{\omega_0} .
\label{correlator_sym}
\end{align}

As a first step towards the construction of a field integral
formulation, we introduce a superfield
$$
\psi \equiv
\left(
  \begin{matrix}
    \psi_\uparrow\cr
    \psi_\downarrow
  \end{matrix}
\right),\qquad \bar \psi\equiv (\bar \psi_\uparrow,\bar \psi_\downarrow),
$$
where component structure refers to the matrix structure in
\eqref{GreenTspace}, and the superfields
$\psi_{\uparrow\downarrow}=\{\psi_{\uparrow\downarrow,\alpha \lambda
  n}\}$ carry complex commuting (anticommuting) components
$\psi_{\alpha=1}$ ($\psi_{\alpha=2}$). The index $\lambda$ will be used to
discriminate between the retarded ($\lambda=+$) and the advanced
($\lambda=-$) sector of the theory. Finally, the commuting variables
${\bar \psi}_{\alpha=1}=\psi^*_{\alpha=1}$ are the complex conjugate
of $\psi_{\alpha=1}$ while in the anticommuting sector ${\bar
  \psi}_{\alpha=2}$ and $\psi_{\alpha=2}$ are independent variables.

We next define the compound Green function
\begin{equation}
G= E^{11}_{\rm AR} \otimes \openone_{\rm BF} \otimes {\tilde G}^+
(\omega_+) + E^{22}_{\rm AR} \otimes \openone_{\rm BF} \otimes {\tilde G}^-
(\omega_-) \,, \label{kernel}
\end{equation}
where matrices with subscript ``AR'' (``BF'') act in the
two-dimensional spaces of $\lambda$ ($\alpha$) indices. Matrices
$E_X^{ij},X=\mathrm{AR,BF}$ contain zeroes everywhere except for a
unity at position $(i,j)$ and $\openone_X$ are unit matrices. The
correlation function $K_\omega$ can then be represented as a
Gaussian integral
\begin{align}
K_\omega (n_1,n_2) &= \Big\langle \int D({\bar \psi},\psi)
\exp\left(-{\bar \psi} G^{-1}\psi\right) \mathcal{F}[\bar \psi_\uparrow,\psi_\uparrow]
\Big\rangle_{\omega_0}\,, \label{functional}
\end{align}
where the pre--exponential factor
$$
\mathcal{F}[\bar \psi,\psi]\equiv {\bar
\psi}_{\uparrow 1+n_2}\psi_{\uparrow 1+n_1}{\bar\psi}_{\uparrow
1-n_1}\psi_{\uparrow 1-n_2}.
$$
The absence of a normalization factor in
\eqref{functional} follows from the fact that the integration over
anticommuting variables normalizes the integral to unity,
$$
\mathcal{Z}\equiv \Big\langle \int D({\bar \psi},\psi)
\exp\left(-{\bar \psi} G^{-1}\psi\right) \Big\rangle_{\omega_0}=1.
$$
We next reformulate the field integral in a way that will later
enable us to make the consequences of the symmetries of the problem
manifest. Beginning with time reversal,  $\mathrm{T}_c$, we use the symmetry of the Green function, $\tilde G=\tilde
G^T$ to write
\begin{align*}
\bar \psi G^{-1} \psi& = {1\over 2}   \left(\bar \psi G^{-1} \psi +
\psi^T
  \sigma^3_{\rm BF} G^{-1} \bar \psi^T\right)\equiv\bar \Psi G^{-1} \Psi,
\end{align*}
where we defined
\begin{equation}
  \label{eq:1}
  \bar \Psi \equiv \frac{1}{\sqrt 2}(\bar \psi,\psi^T
  \sigma^3_\mathrm{BF}),\quad
\Psi \equiv \frac{1}{\sqrt 2}\left(
  \begin{matrix}
    \psi\cr
    \bar \psi^T
  \end{matrix}\right).
\end{equation}
and the Pauli matrix $\sigma^3_\mathrm{BF}$ accounts for the
anti--commutativity of the variables $\psi_{\alpha=2}$, i.e.
$\psi_{\uparrow, 2\lambda n} \psi_{\uparrow, 2\lambda' n'} =
-\psi_{\uparrow, 2\lambda' n'} \psi_{\uparrow, 2\lambda n}$, and the
same for $\psi_\downarrow$. The newly defined integration variables
exhibit the ``reality condition'' (the representative of
$\mathrm{T}_c$ symmetry in the present construction)
\begin{align}
  \label{eq:2}
  \bar \Psi = \Psi^T \tau,
\end{align}
where the matrix $\tau$ is defined through
\begin{align}
  \label{eq:3}
  \tau\equiv E_{\rm BF}^{11}\otimes \sigma_{\rm T}^1+E_{\rm
BF}^{22}\otimes (i\sigma_{\rm T}^2),
\end{align}
and matrices with subscript ``T'' act in the two--component ``time
reversal'' space introduced in \eqref{eq:1}.
% The operation \eqref{eq:1} defines an analogous doubling for the
% constituent fields $\phi$, $\xi$:
% \begin{equation}
%   \label{eq:5}
%   \bar \Psi \equiv (\bar \Phi,\bar \Xi),\quad
% \Psi \equiv \left(
%   \begin{matrix}
%     \Phi\cr
%     \Xi
%   \end{matrix}\right),
% \end{equation}
% where
% \begin{align}
%   \label{eq:6}
% \bar \Phi =\Phi^T \tau,\qquad \bar \Xi=\Xi^T \tau,
% \end{align}
% and the explicit definition of $\Phi$, etc. follows from \eqref{eq:1},
% e.g. $\bar \Phi= \frac{1}{\sqrt 2}(\bar \phi,\phi^T
% \sigma^3_\mathrm{BF})$.
Armed with these definitions, the partition
function may now be rewritten as
\begin{widetext}
\begin{align*}
\mathcal{Z}=\int d\Psi \left\langle e^{-\bar\Psi \Psi- 2 \bar
  \Psi_{\uparrow +} e^{i\left(-\omega_0+\frac{\omega+i\delta}{4}\right)}
  \hat V^+\Psi_{\downarrow +} - 2\bar \Psi_{\downarrow -}e^{i\left(\omega_0+\frac{\omega+i\delta}{4}\right)}\hat V^{-T}\Psi_{\uparrow-}  }\right\rangle_{\omega_0}.
\end{align*}
\end{widetext}

The next step in the construction of the field theory is the average
over the phase $\omega_0$. We will perform this average with the help
of the ``color--flavor'' transformation~\cite{Zirnbauer96},  an exact
integral transform stating that for arbitrary supervectors,
$\Psi_1,\Psi_2,\Psi_{1'},\Psi_{2'}$,
\begin{equation*}
\left \langle e^{\Psi_1^T e^{i\omega_0} \Psi_{2'} + \Psi_{2}^T
    e^{-i\omega_0} \Psi_{1'}}\right\rangle_{\omega_0} =\int d\mu(\tilde Z,Z) \,e^{\Psi_1^T Z\Psi_{1'} + \Psi_{2}^T \tilde
Z \Psi_{2'}},
\end{equation*}
%\begin{align*}
%&\left \langle e^{\Psi_1^T e^{i\omega_0} \Psi_{2'} + \Psi_{2}^T
%    e^{-i\omega_0} \Psi_{1'}}\right\rangle_{\omega_0} \cr
%&\quad=\int d\mu(\tilde Z,Z)
%\,e^{\Psi_1^T Z\Psi_{1'} + \Psi_{2}^T \tilde Z \Psi_{2'}},
%\end{align*}
where integration domain and algebraic structure of the supermatrices
$Z$ and $\tilde Z$ will be discussed in a moment.

Applied to the present context, the transformation obtains
\begin{equation}
\mathcal{Z} = \int d\mu(\tilde Z,Z)\int d\Psi
  e^{-(\bar \Psi_\uparrow \Psi_\uparrow +\bar \Psi_\downarrow \Psi_\downarrow) +2 \bar \Psi_{\uparrow+} \tilde Z  \Psi_{\uparrow-} + 2e^{\frac{i\omega}{2}}\bar \Psi_{\downarrow-}\hat V^{-T} Z\hat
  V^+
     \Psi_{\downarrow +} },
\label{Psi_Z_fun}
\end{equation}
where the supermatrices $Z=\{Z_{ \alpha t n,\alpha't'n'}\}$ are
subject to the (convergence generating) condition $\tilde
Z_{\alpha\alpha} = (-)^{\alpha-1} Z^\dagger_{\alpha\alpha}$. The
integration measure is given by
\begin{align*}
d\mu (Z,\tilde Z) &\equiv d(Z,\tilde Z) {\rm sdet}(1-Z\tilde Z)\cr
&= d(Z,\tilde Z) \exp {\rm str} \ln (1-Z\tilde Z),
\end{align*}
where ``sdet'' and ``str'' are superdeterminant and supertrace,
respectively~\cite{sdetdef}. Notice that these operations include
the angular momentum indices of the theory.  Finally, the
integration domain in the boson--boson sector is restricted by the
constraint $|Z_{11} Z^{\dagger}_{11}|< 1$ (indices in the
BF--sector.)

% Noting that \eqref{eq:6} entails
% \begin{align*}
% \bar \Phi_+ \tilde Z \Phi_- & = \bar \Phi_- \tau^{-1} \tilde Z^T
% \tau \Phi_+ \cr \bar \Xi_- \hat V^{-T} Z \hat V^+ \Xi_+ &= \bar
% \Xi_+ \hat V^{+T} \tau^{-1} Z^T \tau \hat V^- \Xi_-
% \end{align*}
Using Eq. \eqref{eq:2}, the Gaussian integration over variables $\Psi$  may be
performed to obtain
\begin{equation}
  \mathcal{Z}=\int d(Z,\tilde Z)\, e^{\mathrm{str}\,\ln(1-Z\tilde Z)}
  e^{-\frac{1}{2}\mathrm{str}\,\ln(1-\tilde Z \tau^{-1} \tilde
Z^T
  \tau)-\frac{1}{2}\mathrm{str}\,\ln(1-e^{i\omega} \hat U^\dagger Z
  \hat U \tau^{-1} Z^T \tau)}.
\label{eq:4}
\end{equation}
%\begin{align}
%  \label{eq:4}
%  \mathcal{Z}&=\int d(Z,\tilde Z)\, e^{\mathrm{str}\,\ln(1-Z\tilde Z)}\\
%&\times e^{-\frac{1}{2}\mathrm{str}\,\ln(1-\tilde Z \tau^{-1} \tilde Z^T
%  \tau)-\frac{1}{2}\mathrm{str}\,\ln(1-e^{i\omega} \hat U^\dagger Z
%  \hat U \tau^{-1} Z^T \tau)}.\nonumber
%\end{align}
Both the integration measure, and the action of this functional
integral are invariant under transformations
$$Z\to e^{\frac{i \he \hat n^2}{4}} Z
e^{-\frac{i \he \hat n^2}{4}}, \; \tilde Z \to e^{\frac{i \he \hat
n^2}{4}} \tilde Z e^{-\frac{i \he \hat n^2}{4}}.$$
We may use this
freedom to transform the Floquet operator by $e^{\frac{i \he \hat
n^2}{4}}$ to the form
$$
\hat U \equiv \exp\left(\frac{i\he \hat n^2}{2}\right)
\exp\left(\frac{i K\cos \hat \theta}{\he}\right),
$$
which will turn out to be convenient in the following.
To keep the notation simple, we denote the unitarily transformed
operator again by $\hat U$.

So far, all operations have been exact. In the rest of the paper, we
 aim to reduce the functional integral to a more manageable form
describing long--ranged correlations in the system. To this end, we
need to identify field configurations $(Z,\tilde Z)$ of low action.
Field configurations of this type necessarily have to obey the
constraint
\begin{eqnarray}
\tilde Z=\tau^{-1} Z^T \tau. \label{Zsymmetry}
\end{eqnarray}
The reason is that any violation of this symmetry leads to huge
mismatch between the three terms in the exponent of Eq.~(\ref{eq:4})
and thereby to large values of the action.

Let us briefly discuss the conceptual status of the symmetry
relation \eqref{Zsymmetry}. The symmetry essentially involves the
T--degrees of freedom of the theory. Retracing their origin, we
notice that the T--indices have been introduced to account for the
symmetry of the operator (\ref{U}) under transposition -- the
physical $\mathrm{T}_c$--symmetry. The ensuing doubling of indices
of $Z$ fields, and the corresponding symmetry \eqref{Zsymmetry} lead
to the appearance of two distinct low-energy field configurations in
the theory. Formally, these are the field components diagonal and
off-diagonal in T--space, respectively. Physically, these fields
have a status analogous to the diffusion and Cooperon modes in
disordered metals.

The partition function reduced to field configurations obeying
\eqref{Zsymmetry} is given by
\begin{align}
  \label{eq:7}
  \mathcal{Z} &= \int dZ\, e^{-S[Z]},\\
&S[Z]=-\frac{1}{2}\mathrm{str}\,\ln(1-Z \tilde
Z)+\frac{1}{2}\mathrm{str}\,\ln(1-e^{i\omega} \hat U^\dagger Z  \hat U  \tilde Z ).\nonumber
\end{align}

We next turn to the discussion of the second fundamental
symmetry of the problem, the symmetry under $q$--translation in
angular momentum space, $\hat T_q$. The presence of this symmetry
suggests to use a basis of Bloch functions as a preferential basis. A
wave function in angular momentum space would thus be expanded as
\begin{align}
\label{bloch_general}
\psi_n &= \int_b (d\theta) \, e^{i \theta n}
\, \tilde \psi_{n,\theta},\cr
 \tilde \psi_{n,\theta} & =  \sum_{j=-\infty}^\infty e^{-i\theta(qj+n)}  \psi_{n},
\end{align}
where $\int_b (d\theta) \equiv {q\over 2\pi}\int_0^{2\pi/q} d\theta$
and the states $ \psi_{n,\theta}$ are $q$--periodic, $\tilde
\psi_{n+q,\theta} = \tilde \psi_{n,\theta}$. This representation
defines an analogous expansion of the fields $Z$:
\begin{align}
\label{eq:Z_bloch}
  Z_{n,a;n',a'}&=\int_b (d\theta) (d\theta')\, e^{i \theta
    n-i\theta'n'} Z_{n,\theta,a;n',\theta',a'},\cr
Z_{n,\theta,a;n',\theta',a'} &=\sum_{jj'} e^{-i
\theta(qj+n)+i\theta'(qj'+n')} Z_{n,a;n',a'},
\end{align}
where $a\equiv(\alpha,t)$ is a container index comprising BF- and
T-index, and the fields $Z_{n,\theta,a;n',\theta',a'}$ are
$q$-periodic,
$$
Z_{n+q,\theta,a;n',\theta',a'}=Z_{n,\theta,a;n'+q,\theta',a'}
=Z_{n,\theta,a;n',\theta',a'}.
$$
It may be convenient to think of the fields
$Z_{n,\theta;n',\theta'}$ (we suppress indices in the notation
whenever possible) as matrices in a tensor space $\Bbb{R}^q\otimes
\mathcal{H}_\mathrm{B}$, where $\Bbb{R}^q$ is $q$--dimensional reduced
angular momentum space (the 'unit
cell'), and $\mathcal{H}_\mathrm{B}$ is spanned by Bloch angular
wave functions $\psi(\theta)$, also with periodic boundary conditions
$\psi(\theta)=\psi(\theta+2\pi/q)$. Substituting the above expansion
into the action \eqref{eq:7} and using Eq.~\eqref{Tq_sym}, it is
straightforward to verify that the action assumes the form
\begin{align}
  \label{eff_bloch}
  S[Z]=-\frac{1}{2}\mathrm{str}\,\ln(1-Z \tilde
Z)+\frac{1}{2}\mathrm{str}\,\ln(1-e^{i\omega} \hat U_{\hat
  \theta}^\dagger Z  \hat U^{\vphantom{\dagger}}_{\hat \theta}  \tilde Z ),
\end{align}
where the supertrace now extends over the Bloch quantum numbers,
$$
\mathrm{str}(\hat X) = \sum_{n=1}^q \int_b (d\theta)
\,\mathrm{str}(X_{n,\theta;n,\theta}),
$$
and the residual ``str'' traces over the internal indices of $\hat
X$. Further, $\hat U_{\hat \theta}= \{ U(\theta)_{n,n'}
\delta(\theta-\theta')\}$ is an effective Floquet operator, diagonal
in $\mathcal{H}_\mathrm{B}$, with matrix elements
\begin{align}
  \label{floq_eff}
  U_{nn'}(\theta) = e^{i\tilde h {n^2\over 2}}{1\over q}\sum_{k=0}^{q-1} e^{i
    {K\over \tilde h} \cos({2\pi\over q} k + \theta) + i(n'-n){2\pi
      k\over q}}.
\end{align}
It is instructive to compare this with the matrix elements of the
original Floquet operator (defined in unrestricted $n$-space):
\begin{align*}
  U_{nn'}= e^{i\tilde h {n^2\over 2}}\int_0^{2\pi} {d\tilde
  \theta\over 2\pi} \, e^{i
   {K\over \tilde h} \cos(\tilde \theta) + i(n'-n)\tilde \theta}.
\end{align*}
The difference is that the compactification of the theory to a unit
cell $\{0,\dots,q\}$ in $n$--space implies a discretization of the
angular integration, $\int {d\theta\over 2\pi} \to {1\over q}
\sum_k$. Second, the quantized ``momenta'' $2\pi k/q$ get shifted by
the Bloch momentum $\theta \in [0,2\pi/q]$. One may think of
$U(\theta)_{nn'}$ as an effective Floquet operator on a ring (in
$n$--space) of circumference $q$ which is threaded by an
Aharonov--Bohm flux $\theta$. (Of course, this flux is purely
fictitious and does not break time reversal invariance.) It is
straightforward to verify the unitarity of the effective Floquet
operator,
\begin{align*}
  \sum_{n^{\prime\prime}}
  % \int_b (d\theta)
  U(\theta)_{nn^{\prime\prime}} (U(\theta))^\dagger_{n^{\prime\prime}n'}=\delta_{nn'},
\end{align*}
where angular momentum sums $\sum_n\equiv \sum_{n=1}^q$ are now all
meant to run over the unit cell. Finally, notice that the action
\eqref{eff_bloch} for the field basis introduced in \eqref{eq:Z_bloch}
defines a faithful representation of the theory \eqref{eq:7}; there
are no approximations involved.

Let us close this section with a brief discussion of the general
structure of the theory. To this end, we define the
matrix
\begin{eqnarray}
 \label{Q}
Q\equiv g\sigma_{\rm AR}^3 g^{-1}\,, \qquad g\equiv \left(
    \begin{array}{cc}
      1 & Z \\
      \tilde Z & 1 \\
    \end{array}
  \right).\,
\end{eqnarray}
With the formal representation $\hat U \equiv \exp(i\hat G)$, the
zero frequency action can be represented as (cf. Appendix
\ref{sec:derivationSfl})
\begin{eqnarray}
S[Q]|_{\omega=0} = \frac{1}{2}{\rm str}\ln \left[i \tan
\left(\frac{\hat G}{2}\right) Q + 1 \right]. \label{actionfl1}
\end{eqnarray}
This action is
manifestly invariant under transformations $Q \rightarrow g_0^{-1}Q
g_0$ provided that $[g_0,{\hat G}] = 0$ (or, equivalently, $[g_0,
{\hat U}] = 0$). The $g_0$'s fulfilling this condition are given by
\begin{eqnarray}
g_0=\left(
    \begin{array}{cc}
      1 & Z_0 \\
      \tZ_0 & 1 \\
    \end{array}
  \right)\,,
 \label{g}
\end{eqnarray}
where $Z_0\equiv \{Z_{0,\alpha t,\alpha't'} \delta_{nn'}\}$ are unit
matrices in angular momentum space. The zero-mode configurations
\eqref{Q} define the field space of the nonlinear $\sigma$--model
(of orthogonal symmetry) \cite{Efetov97}. The global invariance
under the ``zero-mode'' fluctuations $g_0$ spanning this manifold
bears important consequences for the formulation of the fluctuation
expansion. It implies that we may organize our later analysis of
general fluctuations (fluctuations non-commutative with $\hat U$)
around any zero mode reference configuration $Q_0$. Below, we will
expand around $g_0=0$, or $Q=\sigma_\mathrm{AR}^3$, corresponding to
$Z_0=0$. In the end of the calculation, the extension to generic
$Q$--field configurations may then be obtained by generalization of
the Gaussian fluctuation action to an action displaying the full
$g_0$ rotation invariance.

For later reference, we finally note (cf. Appendix \ref{sec:derivation-eq.-}) that the
 $Q$-matrix representation of the density correlation function
 $K_\omega(n_1,n_2)$ is given by
\begin{equation}
K_\omega(n_1,n_2) =- {1\over 4}\int_b (d\theta_+) (d\theta_-) \,e^{i
    (n_2-n_1)(\theta_+-\theta_-)} \left\langle \mathrm{str}(Q_{n_2\theta_+,n_2\theta_-}
  \,E_\mathrm{AR}^{21}\otimes \mathcal{P})
 \mathrm{str}(Q_{n_1\theta_-,n_1\theta_+}
  \,E_\mathrm{AR}^{12}\otimes\mathcal{P})\right\rangle,
\label{Kn_Q_rep}
\end{equation}
%\begin{align}
%\label{Kn_Q_rep}
% & K_\omega(n_1,n_2) =- {1\over 4}\int_b (d\theta_+) (d\theta_-) \,e^{i
%    (n_2-n_1)(\theta_+-\theta_-)}
%\times\\
%&  \big\langle \mathrm{str}(Q_{n_2\theta_+,n_2\theta_-}
%  \,E_\mathrm{AR}^{21}\otimes \mathcal{P})
% \mathrm{str}(Q_{n_1\theta_-,n_1\theta_+}
%  \,E_\mathrm{AR}^{12}\otimes\mathcal{P})\big\rangle,\nonumber
%\end{align}
where $\mathcal{P}\equiv E^{11}_\mathrm{T} \otimes
E^{11}_\mathrm{BF}$, the average is over the functional with action
\eqref{eff_bloch} and the representation (\ref{Q}) is implied. Note
that the fields $Q$, by definition, are also $q$-periodic, i.e.,
$Q_{n\theta,n\theta'}=Q_{n+q\theta,n+q\theta'}$, and in
Eq.~(\ref{Kn_Q_rep}) $|n_1-n_2|$ may be larger than $q$.

\section{Effective action}
\label{sec:sigma}

In this section we reduce the functional integral formulation
\eqref{eq:7} (or, equivalently, (\ref{eff_bloch})) to a more
manageable field theory of localization in angular momentum space.
In doing so, we need to pay attention to fluctuations $\tilde
Z_{nn'}$ inhomogeneous in angular momentum space. We will begin by
looking at the zero frequency action, $\omega=0$, which describes
the physics of infinitely long--lived correlations.

Conceptually, the field $\tilde Z_{nn'}$ describes the dynamical
evolution of states $|n\rangle \otimes \langle n'|$. (For example,
the action \eqref{eq:7} essentially measures the overlap of the
one--step evolved state, $\hat U \tilde Z \hat U^\dagger
\leftrightarrow \hat U |n\rangle \otimes \langle n'|\hat U^\dagger$
with the un-evolved configuration.) Configurations of low dynamical
action are to be expected if (a) $n$ is close to $n'$ (in which
case, $|n\rangle \otimes \langle n'|$ evolves close to a ``classical
trajectory'' through angular momentum space, and (b) variations in
$n$ are shallow, such that the classical action of the reference
trajectory is low. Indeed, in the limiting case $Z_{nn'} \simeq B_n
\delta_{nn'}$ and $B_n$ independent of $n$, the operators $\hat U$
and $\hat U^\dagger$ in \eqref{eq:7} cancel out and the action
vanishes. Our goal is to compute the action associated to soft
fluctuations around this zero mode limit.

\subsection{Fluctuation action}
\label{sec:fluctuation-action}

As discussed in the end of the foregoing
section, it will be sufficient to analyze angular momentum space
fluctuations around the reference point $Z=0$. The structure of the
full action then follows from its invariance under uniform rotations
$g_0$. We thus start by considering the quadratic expansion of the
action \eqref{eff_bloch} around $Z=0$ and at $\omega=0$. Denoting the
ensuing quadratic action by $S_\mathrm{fl}$, we have
\begin{align}
\label{quadratic_Z}
  S_\mathrm{fl}[Z] &= \frac{1}{2}{\rm str}( \tilde Z Z- \tilde Z {\hat
U}^\dagger Z {\hat U})\\
&={1\over 2}\sum\limits_{nn'n''n'''}\int_b (d\theta) (d\theta')
\,\mathrm{str}(Z_{n\theta,n''\theta'}\tilde
Z_{n'\theta',n'''\theta})
%\cr
%&\hspace{1cm}\times
(\delta_{n,n'''}\delta_{n',n''}-U(\theta')_{n'',n'}U(\theta)^\dagger_{n''',n}).
\nonumber
\end{align}
This is an exact representation of the quadratic fluctuation
action. We now need to identify those configurations $Z$ whose action
is asymptotically vanishing. All other, ``massive'' field
configurations can then be integrated out by perturbation
theory. Following our discussion in the end of the previous section,
we split the fields as
$$
Z=B+C
$$
into ``massless'' configurations $B$ and ``massive'' fluctuations
$C$. The massless sector will be parameterized as
\begin{align}
  \label{B_para}
  B_{n\theta,n'\theta'}=q^{-1} \sum_\phi  e^{-i\phi n}
B_{\theta,\theta'}(\phi) \,\delta_{nn'},
\end{align}
where $\sum_\phi$ is a shorthand for a summation over ``mode
indices'' $\phi=0,2\pi/q,4\pi/q,\dots,\phi_0$ smaller than a certain
cutoff index $\phi_0$ to be discussed in a moment. The fields $B$
describe fluctuations diagonal in angular momentum space, $n=n'$. To
understand the meaning of this limitation, recall the interpretation
of $Z_{n_1n_2}$ as the representative of a bilinear $|n_1\rangle
\langle n_2|$. Turning to a Wigner representation (symbolic
notation), $|n_1\rangle \langle n_2|\to W(n,\theta) \equiv
\sum_{\Delta n} e^{i\Delta n \theta}|n+\Delta n/2\rangle \langle
n-\Delta n/2|$, we notice that the degree of off-diagonality $\Delta
n=n_1-n_2$ encodes information about the direction of propagation in
angular momentum space (as measured by the angular variable $\theta$
of the rotating particle.) On physical grounds, we expect that sense
of direction to decay rapidly if the kicking strength $K$ is
sufficiently large (for fixed $\he$). Indeed we will see in Appendix
\ref{sec:mass-mode-fluct} that the effect of off--diagonal
fluctuations $C_{n,n'}$ amounts to largely innocent $\sim K^{-1}$
corrections to the theory.  The essential information is carried by
the modes $B_{n,n}$ describing fluctuations uniform in angular
space. The Fourier mode decomposition in terms of an index
$\phi<\phi_0$ states that we will concentrate on fluctuations at
scales $n\gtrsim 2\pi/\phi_0$. Deferring the self-consistent
determination of $\phi_0$ to Appendix \ref{sec:mass-mode-fluct} we
here state that $\phi_0 \ll \tilde h/K = \ell^{-1}$, i.e. fields of
low action will fluctuate on scales $\gg \ell$, where the scale
$\ell$ plays a role analogous to that of the elastic mean free path
in disordered systems. It is important that the cutoff index
$\phi_0$ is invariant under the transformation (\ref{eps_class}).
(Note also that the condition $2\pi/q <\phi_0\ll \he/K$ implies
$q\gg \ell$, as we assume above. Otherwise, the massless sector will
contain only the constant mode $B_{n,n}=\mathrm{const.}$)

For the moment we ignore the contribution of the massive modes and
substitute $Z\simeq B$, with $B$ parameterized as in \eqref{B_para}
into the action. This leads to
\begin{align*}
  S_\mathrm{fl}[B]= &{1\over 2} {1\over q^2}\sum_{\phi,\phi'} \int_b
(d\theta)(d\theta')
%\cr
%&\hspace{1cm}\,\times
\mathrm{str}(B_{\theta,\theta'}(\phi) \tilde
B_{\theta',\theta}(\phi')) K_{\theta,\theta'}(\phi,\phi'),
\end{align*}
where the integral kernel is given by
\begin{align*}
  K_{\theta,\theta'}(\phi,\phi')=\sum_{nn'} (\delta_{nn'}-
  U_{nn'}(\theta') U^\dagger_{n'n}(\theta)) e^{-i\phi n-i\phi'n'}.
\end{align*}
Substituting \eqref{floq_eff}, it is a straightforward calculation to
reduce this expression to
\begin{align*}
   K_{\theta,\theta'}(\phi,\phi')&=\left[q-\sum_{k=0}^{q-1}
e^{-i{K\over \tilde h}
     \sin\left( {2\pi k \over q} (\theta-\theta' +
       \phi)\right)}\right]\delta_{\phi,-\phi'} \cr
&
%\hspace{.5cm}
=q\left({K\over 2\tilde h}\right)^2
(\theta-\theta'+\phi)^2\delta_{\phi,-\phi'}+
\mathcal{O}(K\phi/\tilde h)^4.
\end{align*}
In the last expression, we have approximated the discrete $k$--sum
by an integral. (For the small values of $\phi,\theta,\theta'$ under
consideration, this is an innocent operation.) Substituting this
result into the action and switching back to a momentum space
representation $B(\phi) \to B(n)$, we obtain
\begin{align*}
  S_\mathrm{fl}[B]={D_0\over 2} \int_0^q dn\,\mathrm{str}((-i\partial_n
  + [\hat \theta ,\;]) B  (+i\partial_n
  - [\hat \theta ,\;])\tilde B),
\end{align*}
where we have passed from sum to integration because $B(n)$ is
interpreted as a smooth function of the variable $n$. Here, we have
introduced the (bare) diffusion constant
\begin{equation}
  \label{D0}
  D_0=\left({K\over 2\tilde h}\right)^2,
\end{equation}
and the commutators act as
$$
([\hat \theta,B])_{\theta,\theta'}=(\theta-\theta')
  B_{\theta,\theta'}.
$$
The action above was obtained by quadratic expansion around the
reference configuration $Q=\sigma_\mathrm{AR}^3$ (cf. discussion in
the end of Section \ref{sec:functional5}.) To obtain its
rotationally invariant generalization, we rewrite the action as
\begin{align}
  \label{Sfl_gen}
   S_\mathrm{fl}[Q]=-{D_0\over 16} \int_0^q dn\,\mathrm{str}((-i\partial_n
  + [\hat \theta ,\;])  Q  (i\partial_n
  - [\hat \theta ,\;]) Q),
\end{align}
where $ Q = \tilde g \sigma_\mathrm{AR}^3 \tilde g^{-1}$ and
$\tilde g =
\left(\begin{smallmatrix}
  1&B\cr \tilde B &1
\end{smallmatrix}\right)$ and $\tilde g^{-1}
=\left(\begin{smallmatrix}
  1&-B\cr -\tilde B &1
\end{smallmatrix}\right)$ describe infinitesimally small deviations
off $\sigma_\mathrm{AR}^3$. The generalization to an arbitrary
configuration $Q$ is then given by $\tilde g \to g$ and $ Q \to
Q\equiv g\sigma_\mathrm{AR}^3 g^{-1}$, where $g$ need no longer be
close to unity. We thus conclude that action \eqref{Sfl_gen} defined for arbitrary
\begin{align}
  \label{soft_Q}
  Q\equiv g\sigma_{\rm AR}^3 g^{-1}\,, \qquad g\equiv \left(
    \begin{array}{cc}
      1 & B \\
      \tilde B & 1 \\
    \end{array}
  \right),
\end{align}
is the unique rotationally invariant generalization of the quadratic
action above. (Starting from \eqref{eff_bloch}, the same action can be
obtained by the straightforward if more tedious fluctuation
expansion around arbitrary $Z$.)

To conclude the derivation of Eq. \eqref{Sfl_gen}, we need to
explore its stability with respect to the ``massive mode''
fluctuations $C$ that were so far neglected. This analysis is
conceptually straightforward yet tedious in practice. Referring to
Appendix \ref{sec:mass-mode-fluct} for the actual formulation of the
massive mode integration, we here summarize the main conclusions:
\begin{itemize}
\item The soft mode action is stable at length scales $\Delta n
  \gtrsim \ell = {K\over \tilde h}$.
\item Massive mode fluctuations are damped out at scales $<\ell$.
\item Their principal feedback into the soft mode action is a
  weak renormalization of the diffusion constant, $D_0 \to D_q$,
where the so--called quantum diffusion constant
\cite{Izrailev90,Shepelyansky87a} is given by
\begin{equation}
D_q\simeq
D_0(1-2J_2\left(K_q\right)-2J_1^2\left(K_q\right)+2J_3^2\left(K_q\right))+\dots
\label{Deff}
\end{equation}
%\begin{align}
%  \label{Deff}
%  &D_q\simeq \\
%&\;\;D_0(1-2J_2\left(K_q\right)-2J_1^2\left(K_q\right)+2J_3^2\left(K_q\right))+\dots\nonumber
%\end{align}
with
\begin{align}
  \label{Kq_def}
  K_q \equiv {2K\over \tilde h} \sin \left({\tilde h\over 2}\right),
\end{align}
and $J_n(x)$ are  the Bessel functions
of the first kind  (cf. Eq.~\eqref{Bessel_def}.)
\item At certain irrational values~\cite{Casati84}, $\tilde h$ assumes
  the form of an infinite fraction. At these values, long lived
  angular correlations form, and we are not able to get the massive mode
  fluctuations under control. At these values, the theory considered
  in this paper does not apply.
\end{itemize}

\subsection{Frequency action}
\label{sec:frequency-action}

To complete the derivation of the effective action, we need to take
the finiteness of the control parameter $\omega$ into account. This
will generate a second contribution to the action, $S_\omega[Q]$,
adding to the fluctuation action above. In deriving $S_\omega[Q]$,
we will resort to two (parametrically controlled) approximations: first, we will ignore the massive
modes in the computation of the finite frequency action
\cite{weidenmueller85,fn_massive_freq}. Second, we will ignore
operators of $\mathcal{O}(\omega \partial_n^2)$. At large spatial
scales, $\partial_n< \tilde h/K$ and low frequencies, $\omega\ll 1$,
the neglected terms are irrelevant.  Technically, the assumptions above
imply that in the expansion of the action \eqref{eff_bloch} to
lowest order in $\omega$, we will set $Z\simeq B$, and ignore terms
arising from the non-commutativity of $Z$ and $\hat U$ (which would
generate contributions of $\mathcal{O}(\omega \partial_n^2)$.)

We thus start out from
\begin{align*}
  S_\omega[B]&\simeq-\frac{1}{2}\mathrm{str}\,\ln(1-B \tilde
B)+\frac{1}{2}\mathrm{str}\,\ln(1-e^{i\omega} B \tilde B )\cr
&\simeq -{i\omega\over 2} \mathrm{str}\left({B\tilde B\over
1-B\tilde B}\right).
\end{align*}
Comparison with the definition \eqref{soft_Q} shows that this can be
rewritten as
\begin{align*}
  S_\omega[Q]= -{i\omega\over 8} \mathrm{str}\left(Q\sigma_\mathrm{AR}^3\right).
\end{align*}
Combining this with the fluctuation action \eqref{Sfl_gen}, we arrive
at the main result of this paper, the invariant soft mode action,
\begin{widetext}
\begin{align}
  \label{SeffQ}
  S[Q]={1\over 16}\int_0^q dn\, \mathrm{str}\left(D_q (i\partial_n
  - [\hat \theta ,\;])
Q   (i\partial_n
  - [\hat \theta ,\;]) Q -2i\omega
Q\sigma^3_\mathrm{AR}\right).
\end{align}
\end{widetext}
Notice that the action above is universal in that it depends on
system parameters only through the diffusion constant
$D_q=D_q(K,\tilde h)$. The diffusion constant, $D_q$, and therefore
the theory as a whole, are invariant under the transformation $(\he,
K)\to (\he_\epsilon,K_\epsilon)$. (Note that, for $K_\epsilon
\lesssim 1$, the constructions discussed in this paper do not work.)

Conceptually, $S[Q]$ describes the physics of a disordered metallic
\textit{ring} (the fields $Q$ obey periodic boundary conditions
$Q(n) = Q(n+q)$) subject to a fictitious magnetic flux $\theta$.
(Indeed, the action is mathematically similar to the nonlinear
$\sigma$-model action of persistent current system, a disordered
metallic ring threaded by an Aharonov--Bohm flux \cite{Efetov97}). A
magnetic flux through the ring acts on Hilbert space states as
$|n\rangle \langle n'| \to e^{i\theta n} |n \rangle \langle n'|
e^{-i \theta n'}$, and the corresponding ``covariant derivative''
governs the action \eqref{SeffQ}. However, the Aharonov--Bohm
analogy must not be carried too far: the operator $[\hat \theta,\;]$
does not represent a genuine magnetic flux. Rather, it carries the
information on the infinite extension of the ``real'' system, which
is different from the single unit cell system for which
\eqref{SeffQ} is defined. The difference to a real flux is that we
are integrating over all gauge sectors (corresponding to field
configurations $B_{\theta,\theta'}$ for all values $\theta$ and
$\theta'$) at once, and this restores the time reversal invariance
of the system (which in the presence of a real flux would be lost.)
These differences notwithstanding, the Aharonov--Bohm analogy will
be a useful guiding principle in the discussion of the system below.

Let us conclude this section with a general observation that will
simplify our discussion below: in principle, the degrees of freedom
of the theory are ``infinite dimensional matrices'',
$\{B_{\theta,\theta'}\}$ depending on the continuous index $\theta$.
However in the absence of sources such as those specified in
\eqref{Kn_Q_rep}, the integration over this matrix field gives
unity, $\mathcal{Z}=1$. This is a consequence of supersymmetry. In
the presence of a symmetry breaking source (the projector
$\mathcal{P}$ in \eqref{Kn_Q_rep}), supersymmetry gets
\textit{partially} lost. Specifically, the decoupling of the source
from all $\theta$--indices other than $\theta_\pm$ in
\eqref{Kn_Q_rep} implies that the integration over the variables
$B_{\theta,\theta'}$, $(\theta,\theta')\not=(\theta_+,\theta_-)$
does not contribute to the functional integral. These variables may
simply be removed from the integration manifold (as long as we are
probing a functional expectation value with fixed Bloch index
configuration as in \eqref{Kn_Q_rep}.) This reduction is a rigorous
consequence of supersymmetry.  In effect, it implies that the
continuous index set $\{(\theta,\theta')\}$ gets reduced to one
element $(\theta_+,\theta_-)$, i.e. $B_{\theta,\theta'} \to
B_{\theta_+,\theta_-}\delta_{\theta,\theta+}
\delta_{\theta',\theta_-}$.  Since $\theta_\pm$ are fixed, we need
not keep them as indices, i.e. $B_{n\theta_+ a,n\theta_- a'}\to
B_{na,na'}$. Similarly, $Q_{a\theta,a'\theta'}(n)\to Q_{a,a'}(n)$,
where the reduced $Q$ is generated by the reduced $B$. In the
effective action \eqref{SeffQ}, the reduced form of the operator
$\hat \theta$ reads
\begin{align}
\label{theta_red}
  \hat \theta = (\theta_+ E_\mathrm{AR}^{11}+\theta_-
  E_\mathrm{AR}^{22})\otimes \sigma^3_\mathrm{T},
\end{align}
where the matrix $\sigma^3_\mathrm{T}$ represents the sign change in
the Bloch ``flux'' under time reversal. Throughout, we will mostly
work in the reduced representation.

\section{Off resonance}
\label{off resonancs}

In the next two sections, we will employ the effective field theory
\eqref{SeffQ} to discuss the long-ranged physics of the system. Our
discussion is organized in two parts: we first discuss the case of
(near) irrational $\tilde h/4\pi$ where $\xi/q\to 0$ and $\xi \sim
D_q$ will be the localization length of the system. In this regime,
periodicity effects relating to the $q$--commensurability of
Planck's constant play no role and the system can be regarded as
infinitely extended in angular momentum space. In a manner to be
reviewed below, it then becomes similar to an infinitely extended
disordered wire. This limit will serve as a benchmark against which
to compare the physics of the $q$-periodic system.

The commutators $[\hat \theta,\;]$ appearing in the effective action
have an eigenvalue spectrum limited by $2\pi/q$. For $\xi/q\to 0$,
this is negligible in comparison to the excitation gap $\sim \xi^{-1}$
of the localized systems, and we may simply ignore the
$\theta$--contribution to the effective action.

In this regime  the effective action \eqref{SeffQ} reduces to the
nonlinear $\sigma$-model action of quasi one--dimensional disordered
metals~\cite{Efetov97,Efetov83}. To discuss the localization
behavior of this system, we consider the density--density
correlation function defined  in (\ref{correlator}). We first note
that the system is ``translationally invariant'' in angular momentum
space, meaning that $K_\omega(n_1,n_2) = K_\omega(n_1-n_2)$ depends
only on coordinate differences. As such, it can be expressed in a
Fourier representation, $K_\omega(\phi) \equiv \sum_n e^{-i\phi n}
K_\omega(n)$. Comparison with Eq. \eqref{Ediff_Kn} shows that this
function yields the energy increase as
\begin{align}
  \label{Et_Kphi}
  E(\tilde t) = -{1\over 2}\int {d\omega\over 2\pi}e^{-i\omega \tilde
    t} \partial_\phi^2\big|_{\phi=0} K_\omega(\phi).
\end{align}
By unitarity, the function $K$ has to obey
  the limit behavior $\lim_{\phi\to 0} K_\omega(\phi) =
    {i\over \omega}$. The most general low $\phi$ asymptotic
    compatible with this requirement (and time reversal symmetry) reads
  \begin{equation}
    \label{eq:21}
    K_\omega(\phi) = {1\over -i\omega + D(\omega)\phi^2},
  \end{equation}
  where $D(\omega)$ is the so--called dynamical diffusion constant.

Referring for a much more extensive discussion to
Ref.~\cite{Efetov97}, we here briefly discuss the behavior of
$D(\omega)$ upon lowering the frequency:
\begin{itemize}
\item At
  high frequencies, $\omega D_q\to \infty$, fluctuations of the fields
  $Q$ are strongly damped and the effective action
  \eqref{SeffQ} may be approximated by the quadratic expansion around
  $Q=\sigma_\mathrm{AR}^3$:
  \begin{align}
    \label{S_gauss_irr}
     S[B]\simeq {1\over 2}\int d n \, \mathrm{str}(B
  (-D_q\partial_n^2-i\omega ) \tilde B).
  \end{align}
  Likewise, the correlation function \eqref{Kn_Q_rep} reduces to its
  Gaussian representation
  \begin{align}
    \label{Kn_gauss}
    K_\omega(n_1,n_2) = \langle
  \mathrm{str}(B(n_2)\,\mathcal{P})\, \mathrm{str}(\tilde B(n_1)\,\mathcal{P})\rangle.
  \end{align}
  Doing the Gaussian integral, we obtain \eqref{eq:21} with $D(\omega) =
  D_q$ given by its bare value. Substitution into \eqref{Et_Kphi}
  shows that the kinetic energy increases as
\begin{align}
  \label{E_short}
  E(\tilde t) = D_q \tilde t,
\end{align}
i.e. the variable  $E\sim n^2$ shows the behavior characteristic of diffusive dynamics.

\item Upon lowering $\omega$, but still keeping
  $\omega>D_q^{-1}$, corrections to the bare
  value begin to form. The ensuing perturbation series is controlled
  by the parameter $(\omega D_q)^{-1}$. Technically, these corrections may be
  obtained by expansion of the action \eqref{SeffQ} to higher order in
  the constituent fields $B$ and subsequent integration against the
  Gaussian kernel \eqref{S_gauss_irr}. To leading order, this leads to
  a renormalization of the diffusion constant,
  \begin{equation}
    \label{eq:22}
    D(\omega)=D_q\left[1-2\int {d\phi\over 2\pi}
      \frac{1}{-i\omega+D_q \phi^2}\right].
  \end{equation}
  Physically, the correction term represents the modification of the
  diffusion constant by quantum weak localization
  corrections \cite{Altland93,Tian05}. The reduced diffusivity leads to a lowering of the
  energy increase, $E(\tilde t)=D_q \tilde t-\frac{4}{3\sqrt\pi} D_q \tilde
  t^{3/2}$ where $\tilde t\ll D_q$.
\item Finally, at low frequencies, $\omega\ll D_q^{-1}$, the function
$K_\omega(\phi)$ has the form
\begin{eqnarray}
    && K_\omega(\phi)\stackrel{\omega\ll D_q^{-1}}{\longrightarrow}-\frac{A(\phi)}{i\omega},
    \label{eq:28}\\
    && \quad A(\phi)=1-\zeta(3)D_q^2 \phi^2 + {\cal O}(D_q^4\phi^4), \nonumber
  \end{eqnarray}
where $\zeta$ is the Riemann $\zeta$ function. It indicates that the
diffusion constant approaches the value
  \begin{equation}
    \label{eq:23}
    D(\omega) = - i \zeta(3) D_q^2 \omega.
  \end{equation}
  Eq.~(\ref{eq:23}) implies vanishing diffusivity at low frequencies,
  a hallmark of strong dynamical localization.  Substitution into
  \eqref{Et_Kphi} indeed shows that
  \begin{align}
    \label{Et_loc}
    E(\tilde t) \stackrel{t\gg D_q}{ \sim} D_q^2,
  \end{align}
  saturates at a constant value. At the saturation time $\tilde t_L\sim D_q$,
  the two results \eqref{E_short} and \eqref{Et_loc} match.
  Therefore, the localization length scales as $\xi\sim \sqrt{D_q\tilde t_L} = D_q$.

  To make the phenomenon of localization more explicit, one may
  consider the asymptotic form of
  the real time representation of the density correlator, $K(n)\equiv
  K(n,\tilde t\to \infty)$. This function is given by~\cite{Efetov97,Efetov83}
  \begin{eqnarray}
&&   K(n)\stackrel{|n|\gg \xi}{\sim}
      \frac{1}{4\xi}
    \left(\frac{4\xi}{|n-n'|}\right)^{3/2}
\exp\left(-\frac{|n|}{4\xi}\right), \label{eq:24}
  \end{eqnarray}
  where we have neglected the numerical prefactor and have defined
  \begin{equation}
    \label{eq:25}
    \xi=D_q/2
  \end{equation}
  as the
  %dynamical
  localization length. This formula was first obtained by
  Shepelyansky \cite{Shepelyansky87a} based on phenomenological
  arguments.
\end{itemize}

Summarizing, the QKR supports diffusive excitations in angular
momentum space at length scales $\Delta n > K/\tilde h \gg 1$ before
giving way to localization at scales $\Delta n > \xi \sim D_q$.
These results hold for irrational values of $\tilde h$, and
sufficiently large values of the kicking strength, $K
% \gg \tilde h
$. For small values of Planck's constant, $\tilde h\ll1$, the
parameter $K_q$ in \eqref{Kq_def} approaches the classical limit,
$K_q \stackrel{\tilde h \ll 1}{\longrightarrow} K$, and the
diffusion constant becomes
\begin{equation}
  \label{eq:26}
  D_q\stackrel{\tilde h\ll 1}\longrightarrow\left({K^2\over 2\tilde
    h}\right)^2 \left(1-2J_2(K)-2J_1^2(K) + 2 J_3^2(K)\right),
\end{equation}
where the corrections are the classical Rechester-White
corrections~\cite{Rechester80,Fishman00}. In this limit, our result
$\xi=D_q/2$ confirms an empirical formula by Shepelyansky and
co-workers~\cite{Chirikov81,Shepelyansky86,Shepelyansky87,fn_localization}.

\section{Resonances}
\label{resonances}

We now consider the system at rational $\tilde h=4\pi
p/q$~\cite{Casati79}. Pioneering work on the physics of the kicked
rotor at these ``resonant'' values of Planck's constant has been
done by IS \cite{Izrailev79} (see also Ref.~\cite{Izrailev90} for
review.) To assist the orientation of the reader, we begin by
reviewing some key results of that work in the language of the
present formalism. We will then proceed to extend the theory so as
to include the effects of diffusion and Anderson localization.
(While the ramifications of localization at resonant values of
$\tilde h$ have been addressed before \cite{Izrailev79,Chang85},
these discussions were based on phenomenology and numerical
diagonalization.) Throughout this section, it will be convenient to
represent the function $E(\tilde t)$ as (cf. Eq. \eqref{Ediff_Kn})
\begin{align}
  \label{Et_on_res}
  E(\tilde t) =- {1\over 2} \int {d\omega\over 2\pi}\,e^{-i\omega \tilde t}\sum_{n=0}^{q-1} \int_b
    (d\theta) \partial^2_{\theta'}\big|_{\theta'=\theta}K_\omega(n),
\end{align}
where $K_\omega(n)$ is the density correlation function evaluated at
fixed parameters $\theta$, $\theta'$.

\subsection{Theory of Izrailev and Shepelyansky}
\label{sec:theory-izra-shep}

In Ref.~\cite{Izrailev79}, the periodicity effects in angular
momentum space were discussed in terms of an eigenbasis of the
effective Floquet operator \eqref{floq_eff}. To reproduce these
results in the language of the present formalism, we start from the
formal diagonalization
\begin{align*}
  U_{nn'}(\theta) = \sum_j R_{nj}(\theta) e^{i \epsilon_j(\theta)} R^\dagger_{jn'}(\theta),
\end{align*}
where $\hat R(\theta)\equiv \{R_{ij}(\theta)\}$ are unitary $q\times
q$ matrices parametrically depending on the Bloch angle, and
$\epsilon_j(\theta)$ are the $\theta$--dependent eigenenergies. We
may use the transformation $\hat R(\theta)$ to transform the fields
$Z$ in \eqref{quadratic_Z} to an ``eigenbasis representation'',
\begin{align*}
  Z_{n\theta,n'\theta'}\to R_{nj}(\theta) Z_{j\theta,j'\theta'}R^\dagger_{j'n'}(\theta').
\end{align*}
This transformation is unitary and does not affect the integration
measure. As a result the Gaussian expansion of the action assumes the form
\begin{align*}
  S[Z]=&{1\over 2} \sum_{jj'} \int_b (d\theta) (d\theta')
  \,\mathrm{str}(Z_{j\theta,j'\theta'}\tilde Z_{j'\theta',j\theta})
%\cr
%&\quad\times
\left(1-e^{i(\epsilon_{j'}(\theta')-\epsilon_j(\theta) +\omega)}\right),
\end{align*}
where we have included the frequency contribution to the quadratic
action.
Likewise, the Gaussian approximation to the density correlation
function \eqref{Kn_Q_rep} is given by
\begin{align*}
  K_\omega(n_1,n_2) &= \int_b (d\theta) (d\theta') \,e^{i
    (n_2-n_1)(\theta-\theta')}
%\cr
%&\;\;\times
(R_{n_2,j}(\theta) R^\dagger_{n_2,j'}(\theta'))
\,(R_{n_1,j'}(\theta') R^\dagger_{n_1,j}(\theta))
%\cr
%&\;\;\times
\left
\langle\mathrm{str}(Z_{j\theta,j'\theta'}\mathcal{P})\,\mathrm{str}(\tilde
Z_{j' \theta',j\theta}\mathcal{P})\right\rangle\cr &= \int_b
(d\theta) (d\theta') \,e^{i
    (n_2-n_1)(\theta-\theta')}
%\cr
%&\;\;\times
{(R_{n_2,j}(\theta) R^\dagger_{n_2,j'}(\theta'))
\,(R_{n_1,j'}(\theta') R^\dagger_{n_1,j}(\theta))\over
1-e^{i(\epsilon_{j'}(\theta')-\epsilon_j(\theta) +\omega)}},
\end{align*}
where in the last step we did the Gaussian integral over the action
$S[Z]$. This result is but a formal expansion of the density-density
correlation function in a Bloch eigenbasis~\cite{Izrailev79}. It may
be employed to obtain estimates for the energy increase $E(\tilde t)$,
as given by Eq.~\eqref{Ediff_Kn}:
\begin{eqnarray}
  E(\tilde t) = - \frac{1}{2}\sum_{n=0}^{q-1} \int_b
  (d\theta)\partial_{\theta'}^2\big|_{\theta=\theta'}\{(e^{i(\epsilon_{j'}(\theta')-\epsilon_j(\theta))\tilde
t}-1) (R_{n,j}(\theta) R^\dagger_{n,j'}(\theta'))
\,(R_{0,j'}(\theta') R^\dagger_{0,j}(\theta))\}.
\label{E_IS}
\end{eqnarray}
This formula has been derived by IS \cite{Izrailev79}. Except for a
few small values of $q(=1,2,4)$, it cannot be evaluated in closed
terms. However, the general structure of the result indicates that
the energy stored in the system increases as in
Eq.~\eqref{Eincrease}. Specifically, for large times the energy
increases quadratically. To understand this asymptotic behavior,
notice that at large times $\tilde t\gg \Delta^{-1}$, where the
level spacing $\Delta \sim |\epsilon_j - \epsilon_{j+1}|$ is a
measure of the typical separation between neighboring levels, the
dominant contribution to \eqref{E_IS} comes from the diagonal term,
$j=j'$. Describing the auto--correlation of individual levels (or
``bands''), $\epsilon_j(\theta)$,   this term is of ``deep quantum
origin''. In this limit, the $\theta$-derivatives dominantly act on
the rapidly oscillatory phases $\sim \exp(i\epsilon_j(\theta')
\tilde t)$, and we arrive at the estimate
$$
\eta \simeq {1\over 2}\sum_{n=0}^{q-1} \int_b
  (d\theta)
|R_{n,j}(\theta)|^2
\,|R_{0,j}(\theta)|^2 (\partial_\theta \epsilon_j(\theta))^2.
$$
Notice that the sum $\sum_{n=0}^{q-1} \int_b
  (d\theta)
|R_{n,j}(\theta)|^2
\,|R_{0,j}(\theta)|^2$ is (approximately) normalized to unity. This
means that
\begin{align}
\label{eta_formal}
  \eta \simeq {1\over 2} \langle (\partial_\theta (\epsilon_j(\theta)))^2\rangle,
\end{align}
can be interpreted as measure of the typical ``level velocity''
corresponding to the Bloch--phase dependence, and the average is
taken over different quasi-energy levels (for fixed Bloch angle
$\theta$). Corrections to the dominantly quadratic increase arise
from other derivative combinations, where the time derivative acts
once (the linear term in \eqref{Eincrease}) or twice (the term
$b(\tilde t)$) on the diagonalizing matrices.

However, at shorter times $\tilde t<\Delta^{-1}$, inter--band
contributions $j\not=j'$ begin to matter, and these can not be
effectively described by a formal eigenfunction decomposition. In
the next section, we present an analytical theory of resonances in
the regime  $\ell<q<\xi$ which yields the full time dependent
profile $E(\tilde t)$.
% Note that, even at $t\gg \Delta^{-1}$, the mixed
% derivative introduces a linear (in $\tilde t$) correction to the
% quadratic increase. This is indeed the second term in
% Eq.~(\ref{Eincrease}), whose coefficient, $a$, is governed by
% $\partial_\theta (\epsilon_j(\theta))$ and thereby has purely {\it
% quantum} origin. At shorter times,
% % mixed derivatives and
% ``inter-band'' contributions $j\not=j'$, dominated by an oscillatory
% factor $e^{i(\epsilon_{j'}(\theta)-\epsilon_j(\theta))\tilde t}$,
% % (the latter not taken into account in the original paper IS)
% begin to matter, and conspire to produce a dominantly {\it
% classical} linear (in $\tilde t$) scaling (the $b(t)$ term in
% Eq.~(\ref{Eincrease})).
% The results above are formal in that they rely on the unknown
% quasi-energy spectrum/eigenfunctions of the Floquet operator. In the
% next section we present an exact
% % a parametrically controlled
% calculation of $E(\tilde t)$ crossing over from linear to quadratic
% increase for moderately large values $q$ such that $\ell<q<\xi$.
% %where $l\sim K/\tilde h$ and $\xi \sim (K/\tilde h)^2$ are ``mean
% %free path'' and localization length respectively.
In section
\ref{sec:reson-regime-gxi}, we will briefly address the regime of
vary large periodicity intervals, $q>\xi$.

\subsection{Resonances in the regime $\ell<q<\xi$}
\label{sec:reson-regime-lqxi}

We will now address the physics of resonances within the framework
of the theory \eqref{SeffQ}.  Specifically, we will aim to obtain
concrete expressions for the energy increase $E(\tilde t)$
% \eqref{Eincrease}
in terms of universal system parameters.

\subsubsection{Frequencies $\omega\gg \Delta$}
\label{sec:freq-omeg-delta}

We start by considering the regime of frequencies $\omega\gg
\Delta$. Recall that $\Delta = 2\pi/q$ is the effective
quasi--energy level spacing (or the Bloch ``band gap'') of a
periodicity volume. Thanks to the large value of $\omega$, we may
expand $Q$ to quadratic order in $B$ and consider the Gaussian
approximation to the action, \eqref{S_gauss_irr}, where, however,
the derivatives $\partial_n \to
\partial_n + i [\hat \theta,\;]$ have to be generalized to covariant
derivatives and the commutator acts as
\begin{align*}
  [\hat \theta,B]=\theta_+ \sigma_\mathrm{T}^3B-\theta_-
B\sigma_\mathrm{T}^3.
\end{align*}
Switching to a Fourier representation $B(n) = q^{-1} \sum_\phi
e^{-i\phi n} B(n)$ and doing the Gaussian integral against the source terms \eqref{Kn_gauss},
we obtain for $K_\omega(n) \equiv K_\omega(n,0)$
\begin{align*}
  K_\omega(n)= q^{-1} \sum_\phi \int_b (d\theta) (d\theta')
  {e^{-i(\phi+\theta-\theta')n}\over D_q(\phi+\theta-\theta')^2-i\omega}.
\end{align*}
We next substitute this result into Eq. \eqref{Ediff_Kn} and perform
the straightforward integrals over $n$ and $\omega$. As a result, we
obtain Eq.~\eqref{E_short}; the periodicity of the system does not play a
significant role here.

\subsubsection{Frequencies $\omega \lesssim \Delta$}
\label{sec:freq-omeg}

For $\omega \lesssim \Delta$ and flux parameters
$\theta_+\simeq\theta_-$ relevant to the computation of energy
increase, the action of quadratic fluctuations is no longer small in
comparison to unity. This signals a breakdown of perturbation
theory. Indeed, we anticipate from our discussion of Section
\ref{sec:theory-izra-shep} that this is the regime, where the
band--dispersion, $\partial_\theta \epsilon_j(\theta)$, of
individual levels begins to matter. It is known~\cite{Efetov97} that
the fine structure of individual levels of hyperbolic systems is
encoded in non--perturbative (in $\omega$) fluctuations of the field
theory.

Technically, our setup bears similarity to the problem of
``parametric correlations'' \cite{sa93}. There, one usually
considers correlations in the density of states of a system at
slightly different values of some control parameter. In the present
context, the role of that parameter is taken by the Bloch phases
$\theta_\pm$. (For earlier studies of Bloch--phase parameter
correlations applied to the physics of periodically extended
Hamiltonian chaotic systems, see Refs.~\cite{Taniguchi93,Tian09}.)

To set the stage for the non--perturbative calculation, we first
note that for $\omega\lesssim \Delta$, inhomogeneous fluctuations
(modes of finite momentum $\phi$) have a typical action
$E_c/\Delta>1$ (the action of the first non--vanishing
Perron-Frobenius eigenmode of a diffusive system).
% where $E_c$ is the
% inverse of the diffusion time across the periodicity volume.
This is
much larger than the action of the homogeneous mode, $Q(n)\equiv
Q={\rm const.}$, which means that inhomogeneous fluctuations can be
neglected. Substitution of the constant configuration into
\eqref{SeffQ} shows that the ``zero-mode action'' is given by
$$
S[Q]=\frac{\pi}{8\Delta}{\rm str}(D_q[{\hat \theta}, Q]^2-2i\omega
Q\sigma_{\rm AR}^3).
$$

Referring to Ref.~\cite{Efetov97} for an in-depth discussion of the
technical details, we here merely note, that the integration over
$Q$ is best performed in a polar coordinate representation of the
$Q$-matrices. The idea is to represent $Q$ as $Q={\cal R}Q_0{\cal
R}^{-1}$, where $[{\cal R},\sigma_{\rm AR}^3]=0$ and the ``radial''
matrix $Q_0$ is given by
\begin{eqnarray}
Q_0 = \left(
                 \begin{array}{cc}
                   \cos {{\hat \Theta}} & i \sin {{\hat \Theta}} \\
                   -i \sin {{\hat \Theta}} & \cos {{\hat \Theta}} \\
                 \end{array}
               \right)_{\rm AR}, \qquad \qquad \qquad \qquad \qquad \label{parametrization}\\
{\hat \Theta} = \left(
                 \begin{array}{cc}
                   {\hat \theta}_{11} & 0 \\
                   0 & {\hat \theta}_{22} \\
                 \end{array}
               \right)_{\rm BF},\,\, {\hat \theta}_{11} = \left(
                 \begin{array}{cc}
                   \tilde \theta & 0 \\
                   0 & \tilde \theta \\
                 \end{array}
               \right)_{\rm T},\,\,
{\hat \theta}_{22} = i\left(
                 \begin{array}{cc}
                  \tilde \theta_1 & \tilde \theta_2 \\
                  \tilde \theta_2 & \tilde \theta_1 \\
                 \end{array}
               \right)_{\rm T} \qquad \qquad \nonumber
\end{eqnarray}
with $0<\tilde \theta<\pi,\tilde \theta_{1,2}>0$. The action $S[Q]$
is ``rotationally invariant'' in that it couples only to the radial
coordinates $(\lambda\equiv\cos\tilde\theta,
\lambda_{1,2}\equiv\cosh\tilde\theta_{1,2})$ of the polar
decomposition. Of these three variables, $\lambda\in[-1,1]$ is
``compact'', while $\lambda_{1,2}\in [1,\infty]$ parameterize the
non--compact domain of the super integration manifold.

\begin{figure}[h]
  \centering
  \includegraphics[width=8cm]{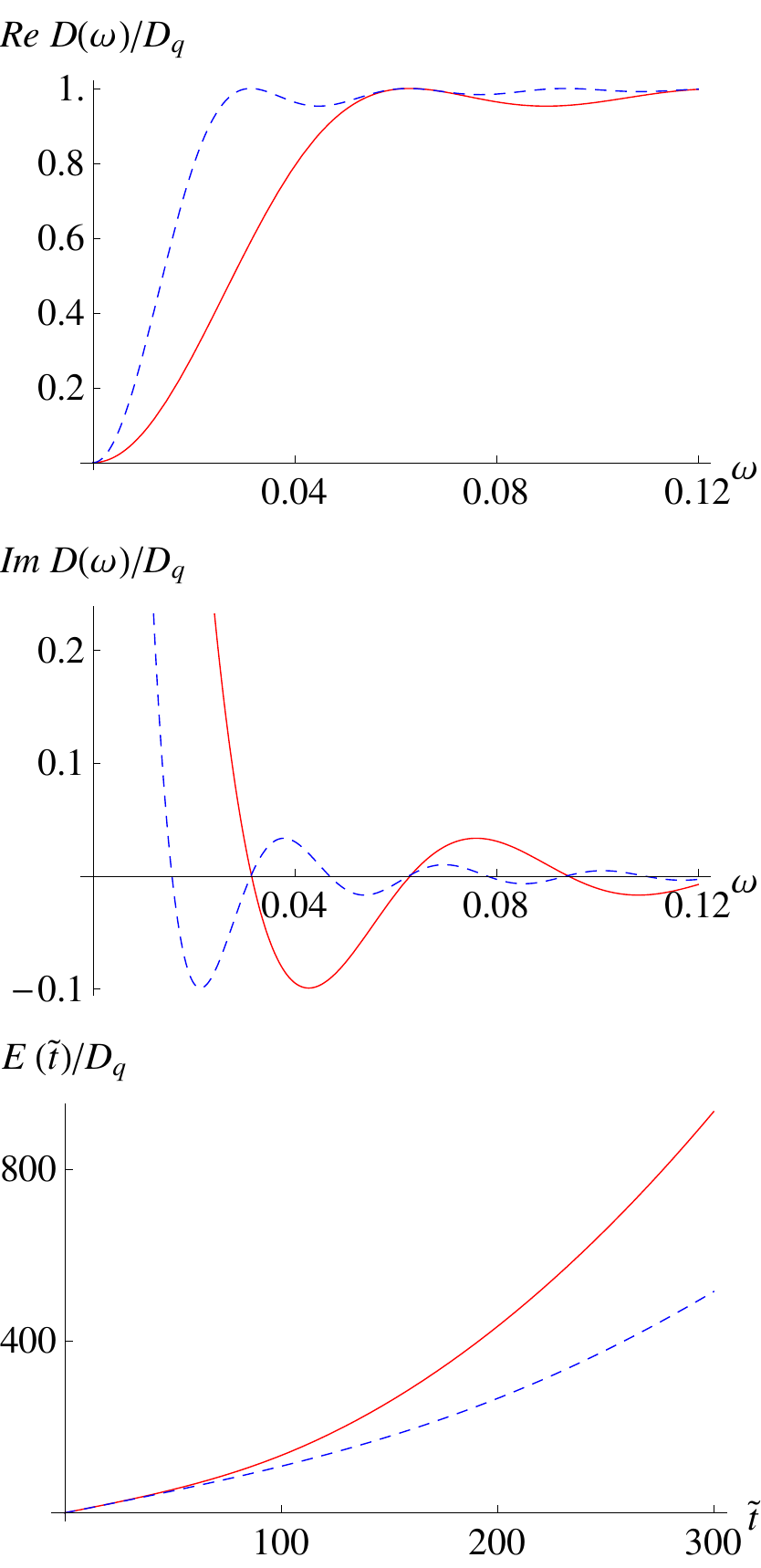}
  \caption{Real (top) and imaginary (middle) part of the dynamical diffusion
  constant at resonant values $\he=4\pi p/q$. Bottom: crossover from linear to quadratic energy increase. The parameters are
  $p=1, q=100$ (solid, in red) and $p=2, q=201$ (dashed, in blue). In
  both cases, $K=5$.}
  \label{crossover}
\end{figure}

Inserting the polar decomposition of $Q$ into the zero-mode action,
we find
\begin{eqnarray}
S = \frac{\pi}{2\Delta}\{D_q[(\Delta\theta)^2
(\lambda_1^2-\lambda^2)+(\theta_++\theta_-)^2(\lambda_2^2-1)] +
2i\omega (\lambda-\lambda_1\lambda_2)\}, \nonumber
\end{eqnarray}
where $\Delta\theta\equiv \theta_+-\theta_-$. The action $S$ then
suggests that, for generic values of the phases $\theta_\pm = {\cal
O}(2\pi/q)={\cal O}(\Delta)$, $Q$--fluctuations off--diagonal in
T-space (characterized by $\lambda_2$) are penalized by a large
action ${\cal O}(D_q\Delta)$. Physically, this means that for fixed
values of the phases $\theta_\pm$ time reversal invariance is
effectively broken (the symmetry gets restored only upon integration
over all phases), and Cooperon--like fluctuations (fluctuations
off-diagonal in T--space) are strongly damped. Under these
circumstances, we may restrict ourselves to T--diagonal fluctuations
(``diffusons'') and set $\lambda_2=1$. The action then collapses to
\begin{eqnarray}
S = \frac{\pi D_q(\Delta\theta)^2}{2\Delta} (\lambda_1^2-\lambda^2)
+ i\frac{\pi \omega}{\Delta} (\lambda-\lambda_1), \label{eq:29}
\end{eqnarray}
and the density correlation function $K_\omega(\Delta\theta)$
% Eq.~\eqref{Kn_Q_rep}
is reduced to
\begin{eqnarray}
K_\omega(\Delta\theta) = \frac{q}{4} \int_1^\infty d \lambda_1
\int_{-1}^1 d\lambda \frac{\lambda_1+\lambda}{\lambda_1-\lambda}
e^{-\frac{\pi D_q(\Delta\theta)^2}{2\Delta} (\lambda_1^2-\lambda^2)
+ i\frac{\pi \omega}{\Delta} (\lambda_1-\lambda)}. \label{integral}
\end{eqnarray}
To proceed further we substitute this result into
Eq.~\eqref{Et_on_res} and perform the remaining integral exactly. As
a result, we obtain
% $$
% E(\tilde t) = \frac{1}{2} \int \frac{d\omega}{2\pi}
% \frac{1-e^{-i\omega\tilde t}}{\omega^2} D(\omega),
% $$
\begin{eqnarray}
E(\tilde t) = -\frac{1}{2} \int \frac{d\omega}{2\pi}
\frac{e^{-i\omega\tilde t}}{\omega^2} D(\omega), \label{eq:30}
\end{eqnarray}
where $D(\omega)$ is given by (see Fig.~\ref{crossover}, the top and
middle panel)
\begin{eqnarray}
D(\omega)= D_q \left[1-\frac{2}{q^2 \omega^2}\left(1-e^{iq
\omega}\right)\right]. \label{eq:36}
\end{eqnarray}
This result for the frequency dependent diffusion constant in
periodic structures was first obtained in Ref.~\cite{Taniguchi93}.
Doing the integral over frequencies in Eq.~(\ref{eq:30}) we obtain
Eq.~\eqref{Et_scaling} (see Fig.~\ref{crossover}, bottom panel).

% \begin{eqnarray}
% E(\tilde t)=D_q q  \left\{ \begin{array}{cl}
%                        {\tilde t\over q} + {1\over 3}\left({\tilde
%                            t\over q}\right)^3,&\quad E_c^{-1}\ll \tilde t\leq q \\
%                        \left({\tilde t\over q}\right)^2 + \frac{1}{3}
%                        , &\quad\tilde t>q.
%                      \end{array}\right.
% \label{eq:34}
% \end{eqnarray}

Eq.~\eqref{Et_scaling} is the main result of the present section. It
accurately describes the function $E(\tilde t)$, up to corrections
in $\exp(-E_c \tilde t)$ (the effect of fluctuations inhomogeneous
across the unit cell). Eq.~\eqref{Et_scaling} coincides (including
the coefficients) with a conjecture on the mean square displacement
of periodic quantum maps (mimicking one-dimensional quantum random
walks) formulated by W$\acute{\rm
  o}$jcik and Dorfman \cite{Dorfman04}. That conjecture was based on
the phenomenological modelling of
% ($q$--dimensional)
quantum maps in terms of random matrix theory. Note that in the
short time regime, $x<1$ or $\tilde t<q$, the quantum correction to
the leading (diffusive) terms scales as $(\tilde t/q)^3$. As pointed
out in Ref.~\cite{Dorfman04}, this scaling reflects universal
fluctuations of the quasi--energy spectra and eigenfunctions of the
system. To the best of our knowledge, this effect has not been
discussed in connection with the QKR.

% such a faster-than-ballistic correction
% (appearing in the $b(\tilde t)$ term of Eq.~(\ref{Eincrease})) is a
% hallmark that fluctuations of quasi-energy spectrum/eigenfunctions
% exhibit certain universal behavior.

Turning to the long time regime, $x>1$ or $\tilde t> q$, comparison with the
formal result (\ref{eta_formal}) leads to the identification
\begin{eqnarray}
\langle (\partial_\theta \epsilon_j (\theta))^2 \rangle =
\frac{2D_q}{q}\approx \frac{K^2}{q \he^2} \label{eq:35}
\end{eqnarray}
of the typical level velocity square. To understand the meaning of
this equation, notice that the so--called Thouless conductance
\cite{Thouless74} of the periodicity volume
$$
g=\frac{E_c}{\Delta}=\frac{1}{\Delta^2}\langle (\partial_\phi
\epsilon_j)^2 \rangle
$$
can be understood in terms of the typical sensitivity of energy
levels to changes in the boundary conditions
$\psi(q)=\psi(0)e^{i\phi}$. In the present context $\phi=q\theta\sim
\theta/\Delta$ is set by the Bloch phase. Substituting this
identification into the formula above and using that $E_c\sim D_q
\Delta^2$, we obtain (\ref{eq:35}). Qualitatively, our results may
be extrapolated into the regime where $\ell$ is comparable to $q$.
Even quantitatively, (\ref{eq:35}) is in good agreement with the
numerical finding of Ref.~\cite{Izrailev79}, where $\langle
(\partial_\theta \epsilon_j (\theta))^2 \rangle $ was found to be $
0.8 K^2/(q \he^2)$.

\subsection{Resonances in the regime $q>\xi$}
\label{sec:reson-regime-gxi}

Let us now briefly explore what happens in regimes where the
periodicity interval exceeds the localization length. Nominally, we
are still dealing with a $q$-periodic quantum system whose spectrum
can be organized in Bloch bands. However, the fact that wave
functions overlap only exponentially weakly
$\mathcal{O}(\exp(-q/(8\xi)))$  across the $q$-volume means that
these bands are exponentially narrow \cite{Izrailev90}. Accordingly,
the level velocities $\partial_\theta \epsilon_j\sim \exp(-q/(8\xi))
$ are exponentially small meaning that the $\mathcal{O}(\tilde t^2)$
increase of the energy in large volumes $q\gg \xi$ will be weighted
by an exponentially small coefficient. We do not aim to analyze this
correction any further.

Turning to the dominant contributions to $E(\tilde t)$, we expect a
profile as in the $q\to \infty$ system,  diffusive spreading at
times $\tilde t<D_q$ followed by saturation at larger times. To see
how this comes about, notice that the principal effect of the
finiteness of $q$ is a substitution $\phi\to \phi + \Delta \theta$
in the argument of the density correlation function
$K_\omega(\phi)$. As far as the function $E(\tilde t)$ is concerned,
this change plays no role, the reason being that the derivative in
\eqref{Et_Kphi} needs to be performed at both $\phi=0$ and $\Delta
\theta=0$. This shows that the function $E(\tilde t)$ behaves as in
the $q\to \infty$ system, apart from the exponentially small tail
corrections mentioned above.

\section{Concluding remarks}
\label{sec:conclusion}

Extending earlier work \cite{Zirnbauer96a}, we have introduced a
microscopic and parametrically controlled theory of quantum
interference and localization in the QKR. Emphasis has been put on
the discussion of resonance phenomena occurring at rational values
$\tilde h = 4\pi p/q$ of Planck's constant. The prototypical theory
\cite{Zirnbauer96a} did not distinguish between rational and
irrational values of $\tilde h/4\pi$, a lack of resolution that has
prompted criticism \cite{Casati98,Casati00}, and led to the
suspicion that the theory was based on hidden ad hoc assumptions.
The present generalization is formulated for rational configurations
from the outset (the irrational limit can be realized in terms of
large coprime values of $q$ and $p$). It is shown that the  system
bears similarity to an effectively disordered ring of circumference
$q$ and subject to an Aharonov-Bohm flux. (The integration over all
values of the flux restores time reversal invariance.) We have
obtained quantitative results for the rotor's kinetic energy
$E(\tilde t)$, an observable that carries detailed information on
its time-dependent transport characteristics. For periodicities $q$
smaller than the localization length $\xi$ (but larger than the
``transport mean free path'' $K/\he$), we have applied
non--perturbative methods to express  the full time dependent
profile $E(\tilde t)$ in terms of the universal scaling form
Eq.~\eqref{Et_scaling}. (Earlier results on $E(\tilde t)$
\cite{Izrailev79} were formulated in terms of expansions of the long
time asymptotics with largely unknown coefficients.)
% The long time asymptotics
% of $E(\tilde t)$ are obtained by non-perturbative methods that
% resolve the flux dependence of individual quasi-energy eigenstates.
For large values $q>\xi$, we have obtained results for the scaling
of $E(\tilde t)$ that are quantitative up to corrections in
$\exp(-q/\xi)$. The resulting profile conforms with general
expectations on energy diffusion in a strongly localized system.

In general, the theory above respects the relevant symmetries of the
rotor (time reversal invariance, $q$-translational invariance, and
the duality to near classical dynamics forming close to fundamental
resonances at $\tilde h = 4\pi p
%+ \epsilon
$). Corrections to the diffusion constant are resolved in agreement
with earlier work
\cite{Shepelyansky86,Shepelyansky87a,Shepelyansky87}. At the lowest
energy scales, the theory resolves the parametric dependence of
individual levels in quantitative agreement with exact quantum
mechanical calculations. In general, we believe that the criticism of
the field theory approach to the rotor \cite{Casati98,Casati00} has
been ill founded: the theory resolves microscopic structures at a
resolution no ``ad hoc'' approach would achieve.

It may be interesting to generalize the approach to other driven
quantum systems, notably the two-sided kicked rotor \cite{Dana95},
the kicked Haper model \cite{Casati92}, the double-kicked particle
\cite{Fishman07}, and the kicked top \cite{Haake88}.

\section*{Acknowledgements}

We are grateful to S. Fishman and I. Guarneri for invaluable
discussions and guidance at various stage of this work. This work
was done in part during the 2007 trisemester ``Statistical physics
out of equilibrium''. C. T. would like to thank the organizer, and
Institut Henri Poincar{\'e} for the hospitality. Work supported by
Transregio SFB 12 of the Deutsche Forschungsgemeinschaft.

\appendix

\section{Derivation of Eq.~(\ref{actionfl1})}
\label{sec:derivationSfl}

For $[W,\sigma_{\rm AR}^3]=0$ we have ${\rm str} \,(W^m) =0$ for odd
$m$. Therefore, $ 2\, {\rm str} \ln (1\pm W) = {\rm str} \ln
(1-W^2)$. Applying this identity to $S[Z]|_{\omega=0}$ (cf.
Eq.~(\ref{eq:7})) we obtain
\begin{eqnarray}
S[Z]|_{\omega=0}= -\frac{1}{2} {\rm str} \Bigg\{\ln \left[1- \left(
                                                   \begin{array}{cc}
                                                     0 & Z \\
                                                     \tZ & 0 \\
                                                   \end{array}
                                                 \right)
\right] -\ln \left[1- \left(\begin{array}{cc}
                                                     {\hat U} & 0 \\
                                                     0 & {\hat U}^\dagger \\
                                                   \end{array}
                                                 \right)
\left(
                                                   \begin{array}{cc}
                                                     0 & Z \\
                                                     \tZ & 0 \\
                                                   \end{array}
                                                 \right)
\right]\Bigg\}. \label{actionfl2}
\end{eqnarray}
Noticing
$$
\frac{1}{2} \left(Q\sigma_{\rm AR}^3+1\right) = \left[1- \left(
                                                   \begin{array}{cc}
                                                     0 & Z \\
                                                     \tZ & 0 \\
                                                   \end{array}
                                                 \right)
\right]^{-1}\,,
$$
we transform Eq.~(\ref{actionfl2}) to
$$
S[Z]|_{\omega=0} = \frac{1}{2} {\rm str} \ln
\left\{\left(\begin{array}{cc}
                                                     \frac{{\hat U}^\dagger - 1}{{\hat U}^\dagger + 1} & 0 \\
                                                     0 & - \frac{{\hat U} - 1}{{\hat U} + 1} \\
                                                   \end{array}
                                                 \right)
                                                 Q +1\right\}.
$$
Upon insertion of ${\hat U}=\exp(-i{\hat G})$ this gets us to
Eq.~(\ref{actionfl1}).

\section{Derivation of Eq.~(\ref{Kn_Q_rep})}
\label{sec:derivation-eq.-}

We first note that the prototype correlation function
\eqref{functional} can be represented as
$$
K_\omega(n_1,n_2) = \partial^2_{\alpha \bar \alpha}\big|_{\alpha=0}
\int D(\bar \psi,\psi)\, e^{-\bar \psi(G^{-1} + X)\psi},
$$
where the source matrix $X$ is defined by
$$
X= E^{\uparrow \uparrow} \otimes E^{11}_\mathrm{BF}\otimes
(\alpha\, E_\mathrm{AR}^{12} \otimes\mathcal{P}^{n_2} + \bar \alpha \,
 E_\mathrm{AR}^{21}\otimes  \mathcal{P}^{n_1}),
$$
where, $\mathcal{P}^n_{n_1,n_2} \equiv
\delta_{n_1,n}\delta_{n_2,n}$ is a projector onto angular momentum
site $n$ and $E^{\uparrow \uparrow}$ projects on the $\uparrow
\uparrow$ sector of the theory. Tracing the fate of this source contribution, we see that
the  action of the functional \eqref{Psi_Z_fun} picks up a contribution
$$
\bar \Psi_{\uparrow}(X\otimes E^{11}_\mathrm{T}+X^T\otimes E_\mathrm{T}^{22}) \Psi_\uparrow.
$$
Further, the first of the logarithms in \eqref{eq:4} gets modified
according to
\begin{align*}
  &\ln(1-\tilde Z \tau^{-1} \tilde Z^T
  \tau)\to\cr
&\;\;\to  \ln\big(1-(\tilde Z+E^{11}_\mathrm{BF}\otimes (\alpha
  E^{11}_\mathrm{T}\otimes \mathcal{P}^{n_2}+\bar \alpha
  E^{22}_\mathrm{T}\otimes \mathcal{P}^{n_1})) \cr
&\hspace{1.5cm}\times  (Z + E^{11}_\mathrm{BF}\otimes (\bar\alpha
  E^{11}_\mathrm{T} \otimes\mathcal{P}^{n_1}+\alpha
  E^{22}_\mathrm{T}\otimes \mathcal{P}^{n_2}))\big),
\end{align*}
where the identification $Z=\tau^{-1} \tilde Z^T \tau$ has been
used. At this point, we may carry out the derivatives with respect
to $\alpha$ and $\bar \alpha$ to obtain
\begin{align*}
  K_\omega(n_1,n_2) = &\big\langle \mathrm{str}((1-\tilde Z Z)^{-1} \tilde Z
(E_\mathrm{BF}^{11} \otimes E^{11}_\mathrm{T}\otimes \mathcal{P}^{n_2})
%\;\cr
%& \times
\mathrm{str}((1- Z
\tilde Z)^{-1}  Z
(E_\mathrm{BF}^{11} \otimes E^{11}_\mathrm{T}\otimes \mathcal{P}^{n_1})\big\rangle,
\end{align*}
where the angular brackets stand for functional averaging over the
functional \eqref{eq:7}, and we have been using the symmetry
\eqref{Zsymmetry}. In the invariant language of $Q$--matrices, this
assumes the form
\begin{align*}
  K_\omega(n_1,n_2) =&- {1\over 4}\big\langle \mathrm{str}(Q
  \,E_\mathrm{AR}^{21}\otimes E_\mathrm{BF}^{11} \otimes E^{11}_\mathrm{T}\otimes
  \mathcal{P}^{n_2})
%\;\cr
%&\quad\times
 \mathrm{str}(Q
  \,E_\mathrm{AR}^{12}\otimes E_\mathrm{BF}^{11} \otimes E^{11}_\mathrm{T}\otimes \mathcal{P}^{n_1})\big\rangle.
\end{align*}
Finally, we may resolve the projector matrices $\mathcal{P}^n$ in the
Bloch basis. Using Eq.~\eqref{bloch_general}, and noting that the
Bloch--momentum is 'conserved', i.e. that the phases
$\theta$--entering the correlation function are pairwise equal, we thus obtain
Eq.~\eqref{Kn_Q_rep}.

\section{Massive mode fluctuations}
\label{sec:mass-mode-fluct}

In Section \ref{sec:sigma}, we have considered the action of
``soft'' fluctuations diagonal and long-ranged in angular momentum
space. We will here extend the analysis of the action to more
general fluctuations. The purpose of this discussion is twofold:
First, we will show that massive mode fluctuations feed back into
the action of soft fluctuations in that they change their
diffusivity. Second, we need to explore the role of massive
fluctuations by way of an a posteriori justification of our
identification of soft modes above.

As we are going to show below, massive fluctuations are limited in
angular momentum space to scales $\Delta n \sim \ell\equiv {K\over
\tilde h} (\gg 1)$. This is smaller than the resonance periodicity
volumes $q>{K\over \tilde h}$ considered in this paper. To keep the
notation simple, we will therefore neglect the Bloch structure of
the fields $Z$ throughout this appendix. In practice, this means
that we will work with the basic representation $Z=\{Z_{ \alpha t
n,\alpha't'n'}\}$ defined in \eqref{Psi_Z_fun}.

\subsection{Massive modes}
\label{sec:slow-modes}

We begin by introducing a few algebraic structures that will
facilitate the analysis below. The fields $\tilde Z=\{\tilde
Z_{nn'}\}$ are matrices in the Hilbert space $\mathcal{H}$ of
angular momentum states $|n\rangle$. As such, they may be viewed as
elements of the Hilbert space ${\mathcal{M}}\equiv
\mathcal{H}\otimes \mathcal{H}^\ast$, where $\mathcal{H}^\ast$ is
the dual angular momentum space, i.e. the space of states $\{\langle
n|\}$. The space ${\mathcal{M}}$ is spanned by the states,
$$
|n,n')\equiv |n\rangle \otimes  \langle n'|.
$$
We will also need the dual of ${\mathcal{M}}$, i.e. the space
${\mathcal{M}}^\ast$ of states
$$
(n,n'| \equiv \langle n| \otimes |n'\rangle.
$$
The natural scalar product and resolution of unity in
${\mathcal{M}}$ are given by, respectively
\begin{align}
  \label{eq:8}
  (n_1, n_1'|n_2, n_2') &= \delta_{n_1,n_2}\,\delta_{n_1',n_2'},\cr
\sum_{nn'}\, |n,n')(n,n'|&=\openone_{{\mathcal{M}}}\,.
\end{align}
Our analysis of fluctuations below will be conveniently performed in
the Fourier transformed basis,
\begin{align}
  \label{eq:9}
  |\phi, \sigma)&\equiv \sum_n e^{-i \phi n} |n,n-\sigma),\cr
  (\phi, \sigma|&\equiv \sum_n \;e^{i\phi n} (n,n-\sigma|,\cr
&\cr |n,n')&= \int_{-\pi}^\pi {d\phi\over 2\pi}\, \;e^{i\phi
n}|\phi,n-n'), \cr (n,n'|&= \int_{-\pi}^\pi {d\phi\over 2\pi}\,
e^{-i\phi n}(\phi ,n-n'|.
\end{align}
Eq.~\eqref{eq:8} implies the orthogonality and completeness relations
\begin{align}
(\phi,\sigma|\phi',\sigma') &=
2\pi\delta(\phi-\phi')\,\delta_{\sigma,\sigma'},\cr \sum_\sigma
\int_{-\pi}^\pi \frac{d\phi}{2\pi}\, |\phi, \sigma)(\phi,
\sigma|&=\openone_{{\mathcal{M}}}\,. \label{basisproperty1}
\end{align}
As discussed in the main text, modes of lowest action have zero
angular momentum difference, $\sigma=0$, and wave numbers
$\phi<\phi_0$, where $\phi_0$ is an upper cutoff to be
self-consistently defined. To make the separation of ``massive'' and
``massless'' modes explicit, we introduce a projection operator $\hat
\pi_0$
\begin{align}
\hat \pi _0  &\equiv  \int_s\frac{d\phi}{2\pi} \, |\phi, 0)( \phi,
0|, \label{projector2}
\end{align}
where $\int_sd\phi \equiv \int_{|\phi|<\phi_0} d\phi$. It is
straightforward to verify that this is a projection, i.e. $\hat
\pi_0^2=\hat \pi_0$. The projection on the complementary sector of
``fast modes'' is given by $\hat \pi=
\openone_{{\mathcal{M}}}-\hat \pi_0$, with the explicit
representation
\begin{align}
\hat \pi &=  \int_f\frac{d\phi}{2\pi} \, |\phi, 0)(\phi, 0| +
\sum_{\sigma\neq 0} \int \frac{d\phi}{2\pi} \, |\phi,\sigma)( \phi,
\sigma | \,, \label{projector1}
\end{align}
where $\int_fd\phi\equiv
\int_{\phi_0\le |\phi|\le \pi}
d\phi$, and $\int d\phi\equiv \int_{-\pi}^\pi
d\phi $. The projection operation defines a decomposition
\begin{eqnarray}
Z= \hat \pi_0 Z+\hat \pi Z \equiv B + C,
\label{separation}
\end{eqnarray}
where $B\equiv \hat \pi_0 Z$ and $C\equiv \hat \pi Z$ are the
massless and massive contribution to $Z$, respectively.

\subsection{Renormalized Gaussian action}
\label{Gaussian}

With these structures in store, we will now re--iterate the
derivation of the Gaussian action. However, our discussion will
extend that of Section \ref{sec:fluctuation-action} in that we will
keep explicit track of $C$-fluctuations. Ignoring contributions
$\mathcal{O}(\omega \partial_n^2)$ (which for the frequencies
$\omega \ll 1$ of physical interest are of no relevance),  the
quadratic action is given by $S[Z]\approx S_{\rm fl}[Z] +
S_\omega[Z]$ with
\begin{align}
S_{\rm fl}[Z] &= \frac{1}{2}{\rm str}( \tilde Z Z- \tilde Z {\hat
U}^\dagger Z {\hat U})\,,
\label{Sfluctuation}\\
S_\omega[Z] &= -\frac{i\omega}{2}\, {\rm str}(
\tilde Z Z)\,. \label{Sfrequency}
\end{align}

\textit{The action $S_\omega$ ---} We first notice that the
projector property of $\hat \pi$ implies a decomposition
$$
S_\omega[Z] = S_\omega[B]+S_\omega[C].
$$
(The proof of this relation is based on
the orthogonality
property
\begin{align}
  \label{trace_orth}
  \mathrm{str}(Z\hat \pi \tilde Z)=\mathrm{str}(Z\hat \pi^2 \tilde Z)=
  \mathrm{str}(\hat \pi Z \hat \pi \tilde Z),
\end{align}
which in turn follows from the definition (\ref{projector2}).) In
principle, the
fast field contribution $S_\omega[C]$ leads to a modification of the
fast fluctuation action to be discussed below. For the small values
of $\omega\lesssim D_0^{-1}$ relevant to our present discussion,
however, this modification is irrelevant. We
therefore dispose of the contribution $S_\omega[C]$ and turn to the slow action
$S_\omega[B]$.
Expanding the fields $B$ as
\begin{equation}
  \label{eq:16}
  B=\int_s \frac{d\phi}{2\pi} B(\phi) |\phi,0),
\end{equation}
where $B(\phi) \equiv (\phi,0|B=(\phi,0|Z$ and using the auxiliary
identity (a direct consequence of Eq.~ (\ref{eq:8}) and the definition
(\ref{eq:9})),
\begin{align*}
  \mathrm{tr}(|\phi,\sigma) Z) = (-\phi,-\sigma|Z) e^{-i \phi \sigma},
\end{align*}
we  obtain
\begin{align*}
  S_\omega[B] \simeq -\frac{i\omega }{2}\int_s {d\phi\over 2\pi}\,
  \mathrm{str}(B(\phi) \tilde B(-\phi)).
\end{align*}
At this point, we may transform back to angular momentum space.
Abbreviating
$$
B(n)\equiv B_{n,n},
$$
and approximating momentum sums by integrations, $\sum_n \to \int dn$,
this gives
\begin{align}
  \label{Somega}
  S_\omega[B] \simeq -\frac{i\omega }{2}\int dn\,
  \mathrm{str}(B(n) \tilde B(n)).
\end{align}

\textit{The action $S_\mathrm{fl}$ ---} We next turn to the more
involved discussion of the fluctuation action. Introducing the
notation $\mathrm{Ad}_{\hat O} X\equiv \hat O^\dagger X \hat O$, the
latter may be written as
\begin{widetext}
  \begin{align}
S_{\rm fl} [B,C]&= \frac{1}{2}{\rm str} \left[(\tilde B +
  \tilde C)
\left(1-\aU\right)(B+C)\right] \cr
&= \frac{1}{2}{\rm str} \big[\tilde B \left(1-\aU\right)B
 -\tilde C
\ppi \aU B - \ppi {\rm Ad}_{\hat U^{-1}}\tilde B C + \tilde C
\left(1-\ppi\aU\right) C \big] \,, \label{flucaction2}
\end{align}
\end{widetext}
where the second line is based on (\ref{trace_orth}).

We next integrate out the massive modes to obtain an effective action
$S_\mathrm{fl}[B]$,
\begin{eqnarray}
e^{-S_{\rm fl}[B]} := \int DC\, e^{-S_{\rm fl}[B,C]} \,. \label{actioneff}
\end{eqnarray}
The Gaussian integration over $C$ can be performed exactly. To
fulfill this task, it is convenient to use the saddle point method.
The solution of the saddle point equation
\begin{align*}
\frac{\delta S_{\rm fl}[B,C]}{\delta C} &= - \ppi {\rm
Ad}_{U^{-1}}\tilde B  + \left(1-\ppi {\rm Ad}_{U^{-1}} \right) \tilde C =
0 \,,
\end{align*}
is given by
$$
\tilde C_{\rm cl} = \left(1-\ppi {\rm Ad}_{\hat U^{-1}}
\right)^{-1} \ppi {\rm Ad}_{\hat U^{-1}}\tilde B .
$$
Applying Eq.~(\ref{Zsymmetry}) to this equation, we obtain
$$
C_{\rm cl} =\left(1-\ppi \aU \right)^{-1} \ppi \aU B.
$$
As usual in supersymmetric theories the integration over quadratic
fluctuations around the saddle point gives a factor of unity, i.e.
the effective action is obtained by straightforward substitution of
the saddle point configurations $(Z^m_\mathrm{cl},\tilde
Z_\mathrm{cl}^m)$ into the action (\ref{flucaction2}). As a result,
we obtain
\begin{align}
% \label{eq:10}
S_{\rm fl}[B] = \frac{1}{2} {\rm str} \left(\tilde B
\left(1-\aU\frac{1}{1-\ppi \aU}\right) B\right),
\nonumber
\end{align}
In the following, it will be convenient to split the action into two contributions,
\begin{align}
\label{eq:11}
S_\mathrm{fl}[B]&=S^0_\mathrm{fl}[B] + S^m_\mathrm{fl}[B],\\
S^0_\mathrm{fl}[B]&=\frac{1}{2} {\rm str} \left(\tilde B
\left(1-\aU\right) B\right),\cr S^m_\mathrm{fl}[B]&=-\frac{1}{2}
{\rm str} \left(\tilde B \left(\aU\frac{\ppi \aU}{1-\ppi \aU}\right)
B\right),\nonumber
\end{align}
where the two terms represent the native slow mode action
($S^0_\mathrm{fl}$) and its renormalization by the inclusion of
massive modes ($S^m_\mathrm{fl}$), respectively. The structure of
the action (\ref{eq:11}) solely  relies on the projector property of
$\hat \pi$, but not on details of the projection operation. We next
aim to bring the action into a more explicit form, and to this end,
we need to take the dynamical structure of the Floquet operator into
account.

Before turning to the calculation, it may be worthwhile to discuss
the physical meaning of the action $S_\mathrm{fl}^m$. This action
describes the renormalization of the soft mode action by ``massive''
fluctuations. Consider a generic massive fluctuation mode
$C_{n_1n_2}$, non-diagonal in angular momentum space. As discussed
in Section \ref{sec:fluctuation-action}, we may loosely interpret
$C_{n_1n_2}$ as the representative of quantum states $|n_1\rangle
\langle n_2|$. Passing to the Wigner representation, the
off-diagonality, i.e., $\Delta n=n_1-n_2\neq 0$, leads to an
excitation in the angular direction of the ``phase space''.
%Representing such states in terms of a Wigner representation
%(symbolic notation) $W(n,\theta) \sim \sum_{\Delta n} e^{i\Delta n
%\theta} |n+\Delta n/2\rangle \langle n-\Delta n/2|$, we notice that
%the degree of off-diagonality, $\Delta n=n_1-n_2$ encodes
%information on the direction of propagation in angular momentum
%space (as measured by the angular variable $\theta$.)
As we will
see, the kernel $(1-\hat \pi \mathrm{Ad}_{\hat U})^{-1}$ describes
the ways in which the memory of initial ``phase space'' information
% $(n,\theta)$
is maintained. For generic irrational values $\tilde
h\ll K$, these effects diminish rapidly. (One may construct certain
infinite fraction representations of $\tilde h$ which sum up to
irrational values and, nonetheless, lead to long-ranged angular
correlations~\cite{Casati84}. At this point, we are not able to get
the ensuing memory kernels under control.) The inclusion of this
memory kernel will lead to a renormalization of the diffusion
constant of the theory. Equally important, it may affect the length
scales at which the dynamics becomes genuinely diffusive.  Again,
the situation bears analogy to the physics of disordered metals.
There, wave packets initially centered around a definite momentum
maintain the memory of their direction of propagation over a certain
time scale, the so-called transport scattering time. In a manner
largely analogous to our present discussion, the transport
scattering time can be seen to determine an effective diffusion
constant.

\textit{The action $S^0_{\rm fl}[B]$ ---} We first consider the
unrenormalized massless action $S^0$. The derivation of this action
essentially repeats the construction of Section
\ref{sec:fluctuation-action} in the present notation and introduces
technical building blocks relevant to the discussion below.
Specifically, we will need the matrix elements of the adjoint
Floquet operator in the two bases $\{|n,n')\}$ and
$\{|\phi,\sigma)\}$. Using that $(n_+,n_-|\mathrm{Ad}_{\hat
  U}|n_+',n_-')=\langle n_+ |\hat U |n_+'\rangle \langle n_-'| \hat
U^\dagger|n_-\rangle$, a straightforward if somewhat tedious
calculation obtains
\begin{widetext}
\begin{align}
  \label{eq:12}
  (n_+,n_-|\mathrm{Ad}_{\hat U}|n_+',n_-')&=i^{n_+-n_+'-n_-+n_-'}
e^{i \frac{\tilde h}{2}(n_+^2-n_-^2)}
 J_{n_+-n_+'}\left(\frac{K}{\tilde h}\right)
J_{n_-'-n_-}\left(-\frac{K}{\tilde h}\right),\cr
(\phi,\sigma|\mathrm{Ad}_{\hat U}|\phi',\sigma') &= 2\pi \tilde
\delta(\phi-\phi'+\tilde h \sigma) e^{-i\frac{\tilde
    h}{2}\sigma^2 + i \frac{\phi'}{2}(\sigma-\sigma')}
J_{\sigma-\sigma'}\left(\frac{2K}{\tilde h}\sin\left(\frac{\phi'}{2}\right)\right).
\end{align}
\end{widetext}
where $\tilde \delta(x)$ stands for that the $\delta$--function
holds only modulo $2\pi$, and
\begin{align}
  \label{Bessel_def}
  J_n(z) = \frac{1}{i^n}\int_0^{2\pi} \frac{d\theta}{2\pi} e^{iz
  \cos\theta} e^{in \theta}.
\end{align}
are the Bessel functions of the first kind. Using this result and
representing the slow fields as (\ref{eq:16}), we then obtain
\begin{align*}
& S^0_\mathrm{fl}[B] = {1\over 2}  \int\limits_s
  \frac{d\phi}{2\pi}\frac{d\phi'}{2\pi}
\mathrm{str}(B(\phi) \tilde B(\phi'))(-\phi,0|1-\mathrm{Ad}_{\hat
  U}|\phi',0).
\end{align*}
Using Eqs. \eqref{basisproperty1} and \eqref{eq:12} this becomes
\begin{equation}
S^0_\mathrm{fl}[B]={1\over 2}  \int_s
  \frac{d\phi}{2\pi}\mathrm{str}(B(\phi) \tilde B(-\phi))
  \left(1-J_0 \left(\frac{2K}{\tilde
h}\sin\left(-\frac{\phi}{2}\right)\right)\right). \label{eq:29}
\end{equation}
%\begin{align}
%\label{eq:29}
%& S^0_\mathrm{fl}[B]\\
%&  ={1\over 2}  \int_s
%  \frac{d\phi}{2\pi}\mathrm{str}(B(\phi) \tilde B(-\phi))
%  \left(1-J_0 \left(\frac{2K}{\tilde
%h}\sin\left(-\frac{\phi}{2}\right)\right)\right).\nonumber
%\end{align}
This action will describe diffusive correlations of the field $B$,
provided the angular kernel $(1-J_0)$ can be approximated as a
\textit{quadratic} function of $\phi$. This is the case, if $\phi$ is
sufficiently small. Specifically, for a soft mode angular cutoff
\begin{equation}
  \label{eq:33}
  \phi_0 \ll \ell^{-1}=\frac{\tilde h}{K}
  % \ll 1
  ,
\end{equation}
with $\tilde h/K$ sufficiently small, we have $\frac{2K}{\tilde
h}|\sin\frac{\phi}{2}|\ll 1$ which means that the Bessel function
may be expanded as
$$
J_0(x) \stackrel{x\ll 1}{=} 1- \frac{x^2}{4}+\mathcal{O}(x^4),
$$
to arrive at $S^0_\mathrm{fl}[Z]\simeq {K^2\over 8\tilde h^2} \int
\frac{d\phi}{2\pi} \phi^2 {\rm str}(B(\phi) \tilde
B(-\phi))$. Finally, transforming back to angular momentum space, we
obtain
\begin{equation}
  \label{eq:14}
   S^0_\mathrm{fl}[B]\simeq {D_0\over 2}\int dn \,\mathrm{str}(\partial_n
   B\partial_n\tilde B).
\end{equation}
Notice that the
condition (\ref{eq:33}) stabilizing this two-derivative action has an
analog in the physics of disordered metals. There, the wave numbers of
diffusive fluctuations (fluctuations governed by a second order
derivative evolution operator) are smaller than the inverse mean free
path, $l_{tr}$, due to impurity scattering. In the present context,
the scale $K/\tilde h$ plays a role analogous to $l_{tr}$ and modes
fluctuating at larger length scales evolve diffusively.

% $$
% Z(n)\equiv Z_{nn}
% $$
% for smooth fields, and approximating momentum sums by integrations, $\sum_n \to \int dn$, we obtain
% \begin{equation}
%   \label{eq:14}
%    S^0_\mathrm{eff}[Z]\simeq {D^0\over 2}\int dn \,\mathrm{str}(\partial_n
%    Z \partial_n\tilde Z),
% \end{equation}
% where the ``diffusion constant''
% \begin{equation}
%   \label{eq:15}
%   D^0\equiv
%   \left(
% \frac{K}{2\tilde h}
%   \right)^2.
% \end{equation}

\textit{The action $S^m_{\rm fl}[B]$ ---} We next discuss the second
contribution $S_\mathrm{fl}^m$ to the fluctuation action
(\ref{eq:11}). Inserting the expansion $B=\int_s {d\phi\over 2\pi}
B(\phi) |\phi,0)$ into Eq.~(\ref{eq:11}) and expanding the resolvent
operator $(1-\hat \pi \mathrm{Ad}_{\hat U})^{-1}$ we obtain our
starting point
\begin{widetext}
$$
S_\mathrm{fl}^m=-{1\over 2} \int_s {d\phi\over
    2\pi}\frac{d\phi'}{2\pi}\,\mathrm{str}(B(-\phi) \tilde B(\phi')) \sum_{n=1}^\infty (\phi,0|\mathrm{Ad}_{\hat U} (\hat
\pi \mathrm{Ad}_{\hat U})^n|\phi',0) .
$$
% \begin{align*}
%  S_\mathrm{eff}^m&=-{1\over 2} \int_s {d\phi\over
%    2\pi}\frac{d\phi'}{2\pi}\,\mathrm{str}(Z(-\phi) \tilde Z(\phi'))\cr
%&\qquad\times \sum_{n=1}^\infty (\phi,0|\mathrm{Ad}_{\hat U} (\hat
%\pi \mathrm{Ad}_{\hat U})^n|\phi',0) .
%\end{align*}
We now insert resolutions of unity (\ref{basisproperty1}) between
the $n$ factors ${\hat \pi}\mathrm{Ad}_{\hat U}$, use that the
projectors $\hat \pi$ appearing are diagonal in the basis
$\{|\phi,\sigma)\}$, i.e. $ \hat \pi |\phi,\sigma) \equiv
\pi(\phi,\sigma)\, |\phi,\sigma), $ where $ \pi(\phi,\sigma) =
\delta_{\sigma,0}\theta(|\phi|-\phi_0) + (1-\delta_{\sigma,0}), $
($\theta(x)$ is the Heaviside function, which is unity for $x\geq 0$
and zero otherwise.) and substitute the explicit form of the matrix
elements (\ref{eq:12}) to arrive at the ``path integral''
representation
  \begin{align}
  \label{eq:32}
    S_\mathrm{fl}^m[Z]&=-{1\over 2} \int_s {d\phi\over
    2\pi}\frac{d\phi'}{2\pi}\,\mathrm{str}(B(-\phi) \tilde B(\phi'))\cr
&\quad\times \sum_{n=1}^\infty \int D(\phi,\sigma)\prod_{i=1}^{n+1}
2\pi \tilde \delta(\phi_i-\phi_{i-1} +\tilde h \sigma_i)
e^{i\sum_{i=1}^{n+1} (-\frac{\tilde h}{2}\sigma_i^2
  +\frac{\phi_i}{2}(\sigma_i-\sigma_{i-1}))} J_{\sigma_i-\sigma_{i-1}}
\left(
\frac{2K}{\tilde h}\sin\left(\frac{\phi_{i-1}}{2}\right)
\right),
  \end{align}
\end{widetext}
Here, the integration measure is given by
$$
\int D(\phi,\sigma) \equiv \prod_{i=1}^n \int_0^{2\pi} d\phi_i \sum_{\sigma_i} \pi(\phi_i,\sigma_i),
$$
and the boundary conditions $(\phi_{n+1},\sigma_{n+1})=(\phi,0)$ and
$(\phi_0,\sigma_0)=(\phi',0)$ are understood. We may now use the
$\tilde \delta$-function constraint in the integral to collapse the
phase factor to multiple $2\pi$,
$$
\sum_{i=1}^{n+1} \left(\tilde h\sigma_i^2
  -\phi_i(\sigma_i-\sigma_{i-1})\right) \equiv 0\, {\rm mod}\, 2\pi.
$$
The constraint further implies
\begin{equation}
  \label{eq:17}
\phi_i \equiv \phi' - \tilde h \sum_{j=1}^i \sigma_j\, {\rm mod}\,
2\pi.
\end{equation}
For generic irrational $\he/(4\pi)$ (for ``non--generic'' values,
see Ref.~\cite{Casati84}), $\phi_i$ is not close to a multiple
$2\pi$ which implies ${2K\over \tilde h}\sin(\phi_i/2)
=\mathcal{O}({K\over \tilde h})\gg 1$. Now for large arguments,
$$
J_\sigma(x)\stackrel{|x|\gg 1}\sim {1\over \sqrt{x}},
$$
which means that Eq.~(\ref{eq:32}) is a controlled expansion in
$(\he/K)^{1/2}\ll 1$. We may truncate this expansion at any desired
order $n$. In conjunction with the near irrationality of $\tilde
h/4\pi$, the finiteness of the sum $\sum_{j=1}^{n+1} \sigma_j\in
\Bbb{N}$ means that $\left(\tilde h \sum_{j=1}^{n+1} \sigma_j\right)
\mbox{mod} \;2\pi$, can not be of
$\mathcal{O}(\phi')<\mathcal{O}(\tilde h/K)$ unless
$\sum_{j=1}^{n+_1} \sigma_j=0$. We thus conclude
\begin{equation}
  \label{eq:18}
\sum_{j=1}^i \sigma_j=0,
\end{equation}
giving $\phi=\phi'$: the integral over intermediate states conserves
the momentum $\phi$.
% (At ``quantum resonances'', the inequality
% (\ref{eq:30}) is violated. Consequently, the present construction
% breaks down, and a modified field theory emerges \cite{at}.)

% Assuming non--resonant values of $\tilde h$,
The considerations above fix the simplified action
\begin{widetext}
  \begin{equation}
    \label{eq:20}
    S_\mathrm{fl}^m[Z]=-{1\over 2} \int_s {d\phi\over
    2\pi}\,\mathrm{str}(B(-\phi) \tilde B(\phi))\sum_{n=1}^\infty \sum_{\{\sigma\}}
\prod_{i=1}^{n+1}J_{\sigma_i-\sigma_{i-1}} \left( \frac{2K}{\tilde
h}\sin\left(\frac{\phi_{i-1}}{2}\right) \right),
\end{equation}
%  \begin{align}
%    \label{eq:20}
%    S_\mathrm{eff}^m[Z]&=-{1\over 2} \int_s {d\phi\over
%    2\pi}\,\mathrm{str}(Z(-\phi) \tilde Z(\phi))\\
%&\quad\times
%\sum_{n=1}^\infty \sum_{\{\sigma\}} \prod_{i=1}^{n+1}J_{\sigma_i-\sigma_{i-1}}
%\left(\frac{2K}{\tilde h}\sin\left(\frac{\phi_{i-1}}{2}\right)\right),\nonumber
%  \end{align}
where $\phi_i$ is given by (\ref{eq:17}) and
$\sum_{\{\sigma\}}=\prod_i \sum_{\sigma_i}$, subject to the constraint
(\ref{eq:18}).

\begin{figure}[h]
  \centering
  \includegraphics[width=17cm]{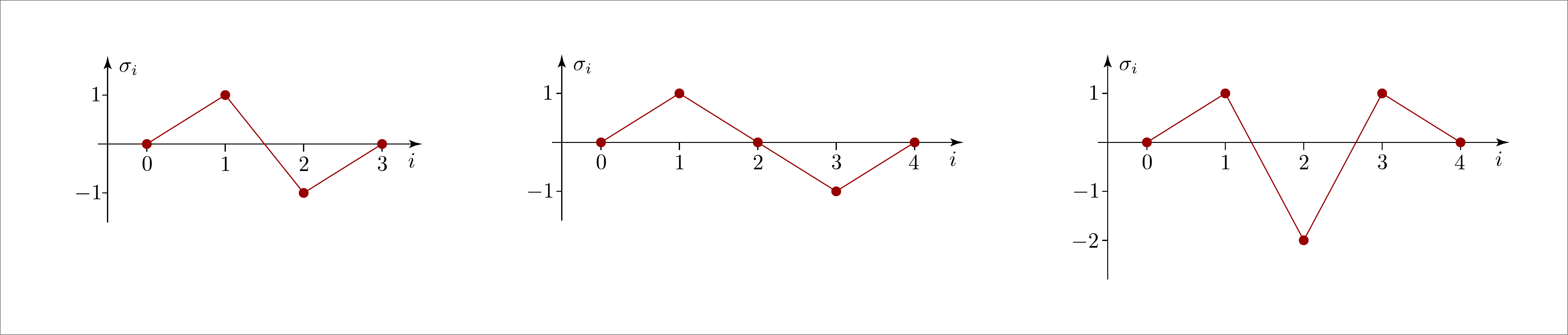}
  \caption{\label{fig:path_sum} Path of lowest order contributing to the renormalization of the diffusion constant.}
\end{figure}

It is instructive, to think of the set $\{(\phi_i,\sigma_i)\}$ in
terms of a path
\begin{align}
  \label{eq:19}
%&
(\phi,0)\to  (\phi-\tilde h
  \sigma_1,\sigma_1)\to
 (\phi-\tilde h
  \sigma_1-\tilde h \sigma_2,\sigma_2)\to \dots
%  \to \cr &\;
\to  (\phi-\tilde h
  \sigma_1-\dots -\tilde h \sigma_{n-1},\sigma_{n-1})\to
(\phi, \sigma_n)\to (\phi,0).
\end{align}
\end{widetext}
\noindent We may imagine the $n+1$ links ``$\to$'' of this path to
be weighted by Bessel functions. Specifically, the first and last
link, respectively carry weights $J_{\sigma_1}\left({2K\over \tilde
h}
  \sin\left(\frac{\phi}{2}\right)\right)$ and $J_{-\sigma_n}\left({2K\over \tilde h}
  \sin\left(\frac{\phi}{2}\right)\right)$. These weights imply a further effective
constraint on the sum over configurations $\{\sigma\}$: since
$\phi<\phi_0\ll \tilde h/K$, the terminating Bessel functions carry
arguments $\frac{2K}{\tilde h} |\sin\frac{\phi}{2}|\simeq |K
\phi/\tilde h|\ll 1$. We may thus apply the asymptotic formula
$$
J_\sigma(\epsilon) \stackrel{|\epsilon|\ll 1}{=}
{\epsilon^{|\sigma|}\over
  2^{|\sigma|} \sigma!}+\dots,
$$
to conclude that the dominant contribution to the sum has
\begin{equation}
  \label{eq:sigma_1n}
  |\sigma_1|=|\sigma_n|=1.
\end{equation}
The terminal weights then yield a factor
\begin{align*}
&  J_{\sigma_1}\left({2K\over \tilde h}
    \sin\left({\phi\over 2}\right)\right) J_{-\sigma_n}\left({2K\over \tilde h}
    \sin\left({\phi\over 2}\right)\right)
%\cr
%&\qquad
\simeq-\left({K \phi \over
      2\tilde h}\right)^2\mathrm{sgn}(\sigma_1 \sigma_n).
\end{align*}
% The phases $\phi_{i-1}$ of all other Bessel function weights are of
% ${\cal O}(\tilde h/K)$ (cf. Eq. (\ref{eq:17}).)
% Regardless of the value of $\tilde h$, this implies that the
% arguments of the inner Bessel functions exceed unity. (For $\tilde h \ll K$, we have $2K \sin(\phi_i/2)/\tilde h \gg (2K/\tilde h) \sin
% (\tilde h/(2K )) \sim 1$.)
%(For
%$\tilde h \ll 1$, we have $2K \sin(\phi_i/2)/\tilde h =
%\mathcal{O}((2K/\tilde h) \sin (\tilde h )) \sim K\gg 1$ and for
%$\tilde h =\mathcal{O}(1)$, $(2K/\tilde h) \sin(\phi_i/2) =
%(2K/\tilde h)\mathcal{O}(1)\sim K/\tilde h \gg 1$.)
For the arguments of the inner Bessel functions, regardless of the
value of $\he$, they much exceed unity, which means that the sum
above will be dominated by the shortest paths obeying the criteria
(\ref{eq:18}) and (\ref{eq:sigma_1n}).

The contributions of leading and sub--leading order to the path sum
are shown in Fig.~\ref{fig:path_sum}. Evaluating these paths in
(\ref{eq:20}), we obtain
 \begin{align}
    \label{Sm_gen}
    S_\mathrm{fl}^m[B]&={D_0\over 2} \int_s {d\phi\over
    2\pi}\,\phi^2\,\mathrm{str}(B(-\phi) \tilde B(\phi))
%\cr
%&\times
\left(-2J_2(K_q)-2J_1^2(K_q)+2J_3^2(K_q)\right).
\end{align}
Passing back to angular momentum space and adding to
$S_\mathrm{fl}^m$ the compounds (\ref{eq:14}) and (\ref{Somega}), we
arrive at the final form of the Gaussian action
\begin{align*}
  S[B] ={1\over 2}\int dn \, \mathrm{str}(B(n)
  (-D_q\partial_n^2 - i \omega) \tilde B(n)),
\end{align*}
where the diffusion constant is given by \eqref{Deff}. Corrections
to the above diffusion constant can be computed by including path of
higher complexity and will be of $\mathcal{O}(K_q^{-3/2})$. However,
at this order in the expansion, anharmonic fluctuations of
$\mathcal{O}(\mathrm{str}(B\tilde C B \tilde C))$ begin to play a
role and the computation becomes significantly more complicated (see
Ref.~\cite{Shepelyansky87a} for the form of the ensuing
corrections.)


\begin{thebibliography}{}
\bibitem{sdetdef} These operations are defined as $\mathrm{str}(\hat
  A)=\mathrm{tr}(\hat A_{11})-\mathrm{tr}(A_{22})$ and
  $\mathrm{sdet}(\hat A)=\mathrm{det}(\hat A_{11}-\hat A_{12} \hat
  A_{22} \hat A_{21})/\mathrm{det}(\hat A_{22})$, respectively, where the
  indices distinguish between commuting and anti--commuting variables.
\bibitem{fn_massive_freq} For the low frequencies of interest to our theory
$\omega\ll 1$, the action of the massive modes (discussed in
Appendix \ref{sec:mass-mode-fluct}) is only insignificantly affected
by $\omega$. Conversely, the coupling constant governing the
frequency action of the soft modes is protected by a Ward identity
\cite{weidenmueller85} and is not changed by massive mode
fluctuations.
\bibitem{fn_localization} In the limit $\he \ll 1$, the density
correlator (\ref{eq:24}) gives $K(n)\sim
\exp\left(-\frac{2|n|}{(K/\he)^2}\right)$ for $|n|\gg K/\he$. This
expression agrees qualitatively with the numerical experiments
\cite{Shepelyansky86} but quantitatively not. There, it was found
that, if one initially excites one level (with zero angular
momentum), the steady-state distribution probability over
unperturbed levels is reached after long times, read $\sim
\exp\left(-\frac{4|n|}{(K/\he)^2}\right)$. This is much smaller than
our analytical results.
\end{thebibliography}
\end{document}